%% file: paper.tex
\documentclass[12pt]{iopart}
\usepackage{iopams}
\usepackage{setstack}    

\usepackage{graphicx}
\usepackage{subfig}
\usepackage{subeqn}
\usepackage{cite}

\begin{document}

\title[Dynamical nuclear spin polarization induced by electronic current through DQDs]{Dynamical nuclear spin polarization induced by electronic current through double quantum dots}

\author{Carlos L\'opez-Mon\'is$^1$, Jes\'us I\~narrea$^2$ and Gloria Platero$^1$}
\address{$^1$Instituto de Ciencia de Materiales de Madrid, CSIC, Cantoblanco, Madrid 28049, Spain.}
\address{$^2$Escuela Polit\'ecnica Superior, Universidad Carlos III, Legan\'es, Madrid 28911, Spain.}

\begin{abstract}
We analyze electron spin relaxation in electronic transport through coherently coupled double quantum dots in the spin blockade regime. In particular, we focus on hyperfine interaction as the spin relaxation mechanism. We pay special attention to the effect of the dynamical nuclear spin polarization induced by the electronic current on the nuclear environment. We discuss the behaviour of the electronic current and the induced nuclear spin polarization versus an external magnetic field for different hyperfine coupling intensities and interdot tunnelling strengths. We take into account, for each magnetic field, all hyperfine mediated spin relaxation processes coming from the different opposite spin levels approaches. We find that the current as a function of the external magnetic field shows a peak or a dip, and that the transition from a current dip to a current peak behaviour is obtained by decreasing the hyperfine coupling or by increasing the interdot tunnelling strength. We give a physical picture in terms of the interplay between the electrons tunnelling out of the double quantum dot and the spin flip processes due to the nuclear environment.
\end{abstract}


\submitto{\NJP}

\maketitle

\input{introduction.tex}

\input{model.tex}

\input{results.tex}

\input{conclusion.tex}

\ack
We are grateful to A. M. Lunde, C. Emary and L. L. Bonilla for enlightening discussions. We acknowledge financial support through Grant No. MAT2008-02626 (MICINN), from FPU grant (C. L\'opez-Mon\'is) and from ITN under Grant No. 234970 (EU).

\section*{References}

\bibliography{mybib}{}
\bibliographystyle{phcpc}%nar

\end{document}

%% file: introduction.tex
\section{Introduction}

In the last decade solid-state spintronics and quantum computing have experienced a great development. In particular, quantum dots (solid-state fabricated zero dimensional devices) have been widely investigated both experimentally and theoretically. For quantum computing and quantum information they have become major candidates for implementing quantum bit units~\cite{Loss_PRA_1998} ({\it qubits}), but also from a fundamental point of view, since quantum dots resemble artificial atoms, they are highly interesting systems for studying basic atomic physics. In this context, spin decoherence and relaxation are among the most desirable mechanisms to be understood since they represent the main sources of quantum computing errors.

%experimentally~\cite{Fujisawa_Science_1998,Ono_Science_2002,Huttel_PRB_2004,Ono_PRL_2004,Rogge_APL_2004,Johnson_PRB_2005,Petta_Science_2005,Liu_PRB_2005,
%Koppens_Science_2005,Pfund_PRL_2007,Koppens_PRL_2007,Baugh_PRL_2007,Petta_PRL_2008,Buitelaar_PRB_2008,Shaji_Nature_2008,Reilly_Science_2008,Pfund_PRB_2009,Churchill_Nature_2009,
%Danon_PRL_2009,Gullans_PRL_2010} and theoretically~\cite{Khaetskii_PRB_2000,Erlingsson_PRB_2001,Merkulov_PRB_2002,Taylor_PRL_2003,Schliemann_JPhys_2003,Eto_JPSJ_2004,Coish_PRB_2004,Erlingsson_PRB_2005,Coish_PRB_2005,Deng_PRB_2005,Coish_PRB_2005,
%Jouravlev_PRL_2006,Coish_JAP_2007,Rudner_PRL_2007,Ramon_PRB_2007,Qassemi_PRL_2009,Fischer_PRB_2009,
%Palyi_PRB_2009,Stopa_PRB_2010,Erbe_PRB_2010,Rudner_PRB_2010}

We investigate spin relaxation in double quantum dots (DQD). Spin Blockade (SB)~\cite{Ono_Science_2002} is a very suitable regime for attempting single electron spin manipulation because two electrons are trapped in a DQD, since Pauli Exclusion Principle avoids electron transport through the dots. Moreover, SB is attainable through transport experiments in DQDs in several materials.
%~\cite{Ono_Science_2002,Koppens_PRL_2007,Pfund_PRL_2007,Shaji_Nature_2008,Churchill_Nature_2009}
However, spin relaxation processes can partially destroy the SB releasing the trapped electrons, leading to a small, though still measurable leakage current ($\sim$ pA-fA)~\cite{Ono_Science_2002,Koppens_Science_2005,Pfund_PRL_2007,Shaji_Nature_2008,Churchill_Nature_2009}. Spin-orbit coupling~\cite{Khaetskii_PRB_2000}, cotunnelling~\cite{Johnson_PRB_2005,Vorontsov_PRL_2008} and hyperfine (HF) interaction between the DQD electrons and the surrounding nuclei spins of the host material represent the main mechanisms for spin-relaxation.
%~\cite{Koppens_PRL_2007,Pfund_PRL_2007,Churchill_Nature_2009} 
Depending on the material, one or more mechanisms may be involved collaborating or competing~\cite{Koppens_Science_2005,Pfund_PRL_2007,Churchill_Nature_2009}.

\begin{figure}
  \centering \subfloat[DQD, leads and the lattice nuclei.]{\label{dqd}\includegraphics[width=0.4\textwidth]{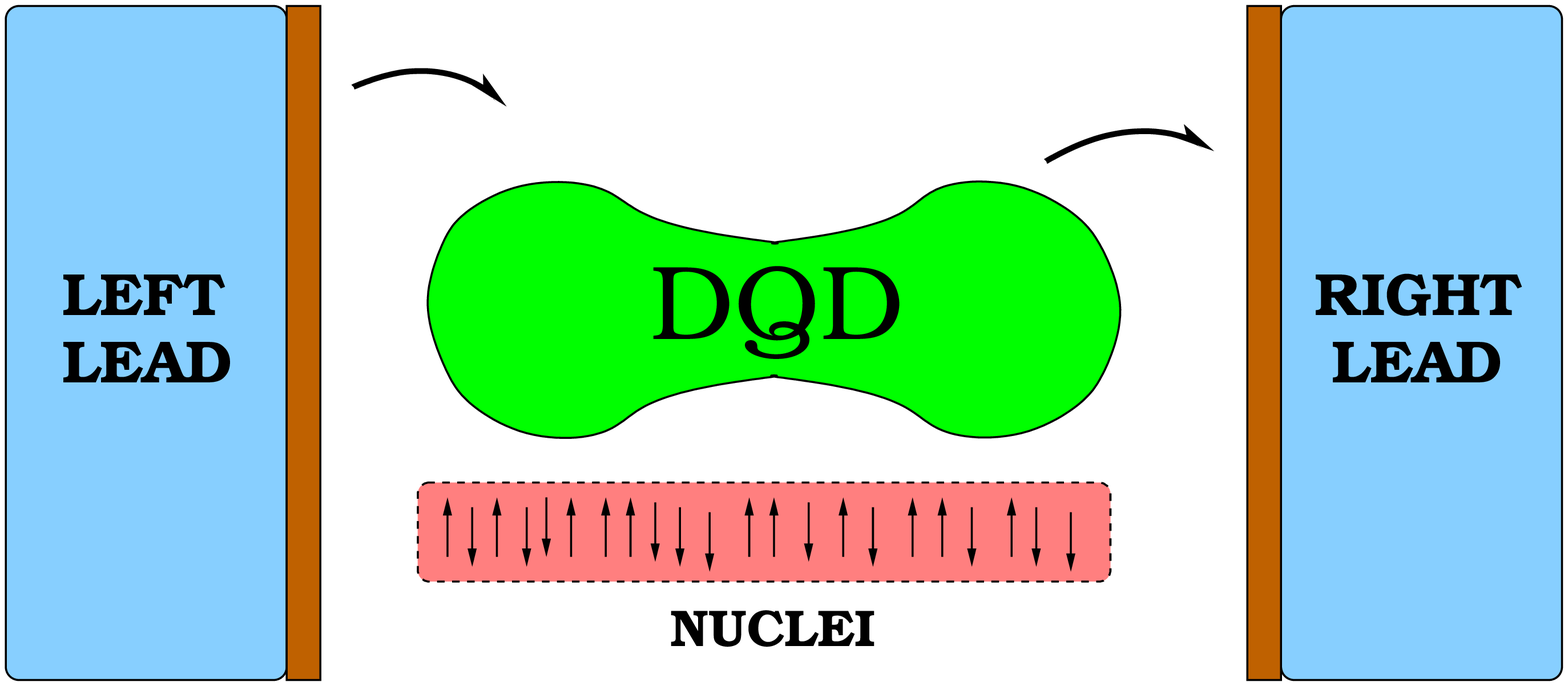}} \hspace{0.2cm}
  \centering \subfloat[Transport scheme.]{\label{sqm}\includegraphics[width=0.43\textwidth]{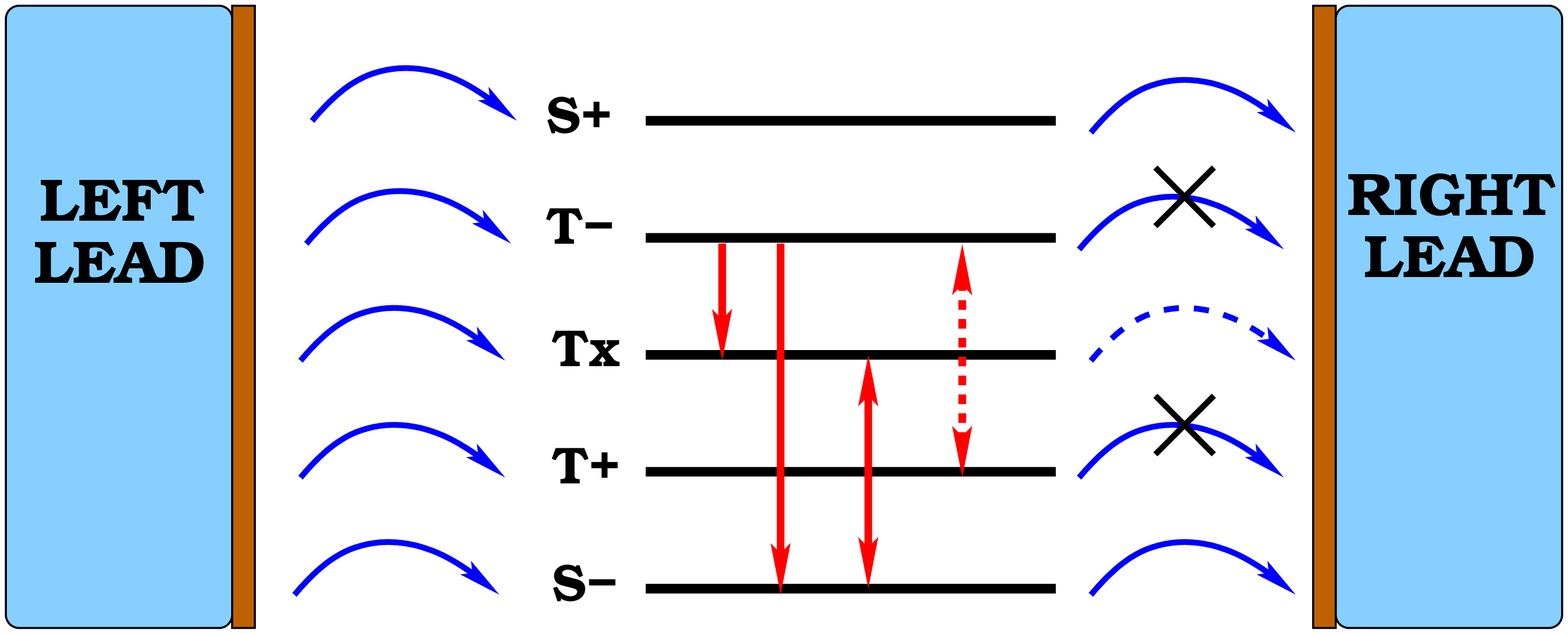}}
  \caption{(Colour online.) a) A coherently coupled DQD is coupled through tunnelling barriers to leads and to the surrounding nuclei of the host material. b) Transport scheme. Electrons tunnel from the left lead into the DQD states (solid blue arrows online). When electrons fall into the singlet states ($S_{\pm}$), tunnelling out of the DQD to the right lead is allowed (solid blue arrows online). However, when electrons fall into a triplet state ($T_{\pm}$) Pauli Exclusion principle prevents them from tunnelling out of the DQD into the right lead, thus, SB occurs. Finally, if electrons tunnel into the $T_x$ state (a mixture of a singlet state and the antiparallel spins triplet $T_0$), they can only tunnel out of the DQD to the right lead if there is a net nuclear spin polarization (dashed blue arrow online), otherwise $T_x$ becomes $T_0$, which is also a blocked state. Electrons trapped in $T_{\pm}$ states interact with the nuclei and relax to a singlet state or to the $T_x$ state (solid red arrows online) through spin-flip processes. However, electrons in $T_x$ state can also spin-flip back to a SB state (red dashed arrows online). Therefore, when electrons are in the $T_x$ state there is a competition between the tunnelling rate from the $T_x$ to the right lead, and the spin-flip rate from $T_x$ to the $T_{\pm}$ triplets (a detailed discussion is given in \sref{mdl}). We will show that this competition will give rise to different physical features in the tunnelling current as a function of an external magnetic field (see~\sref{rslts}).}
  \label{dqdsqm}
\end{figure}

In this paper, we study theoretically spin relaxation in a DQD in SB regime due to HF interaction with the lattice nuclei spins, which has been in the last years a very active field both experimentally~\cite{Koppens_Science_2005,Pfund_PRL_2007,Churchill_Nature_2009,Ono_PRL_2004,Petta_Science_2005,Koppens_PRL_2007,Baugh_PRL_2007,Petta_PRL_2008,Reilly_Science_2008,
Danon_PRL_2009,Gullans_PRL_2010} and theoretically~\cite{Eto_JPSJ_2004,Jouravlev_PRL_2006,Rudner_PRL_2007,Qassemi_PRL_2009,Inarrea_PRB_2007,Dominguez_PRB_2009}. We pay special attention to the dynamical nuclear spin polarization {\it induced by the electronic leakage current}~\cite{Baugh_PRL_2007,Eto_JPSJ_2004,Inarrea_PRB_2007,Inarrea_APL_2007,Inarrea_APL_2009} emerging from the spin-relaxation transitions (\fref{dqdsqm}). This effect is often not taken into account when studying the current through the DQD in SB regime. 
%Nuclear spin polarization has been studied to some extent considering the DQD coupled only to the nuclei spins~\cite{Petta_Science_2005,Rudner_PRL_2007,Gullans_PRL_2010}. However, current has not been sufficiently investigated taking dynamical nuclear polarization into account. 

In the present work we calculate both the electronic leakage current and the nuclei spin polarization induced by the electrons tunnelling through the DQD. In the SB regime, two electrons in the DQD can be either in a triplet or in a singlet state. Nevertheless, current is only allowed to pass through the DQD when they are in a singlet, otherwise they remain trapped in a triplet state. HF interaction mixes triplet and singlet subspaces, and thereby lifts SB. The mixing is due to the different HF interaction strengths, i.e., two different effective magnetic fields (induced Overhauser fields on the electrons by the nuclei), within each dot. In addition, scattering processes between electron and nuclei spins that lead to spin relaxation, also induce a non-negligible nuclear spin polarization as the current flows through the DQD. Moreover, this induced nuclear spin polarization itself, as we shall see below, modifies the mixing between the triplet and singlet subspaces, acting back on the electronic current through the DQD. Furthermore, we find that the Overhauser field proportional to the electron current induced nuclear spin polarization, is in general larger than the one obtained when just considering HF interaction as a random stationary magnetic field
%~\cite{Jouravlev_PRL_2006} 
acting on the electron spins.
%\cite{Petta_Science_2005,Rudner_PRL_2007,Danon_PRL_2009,Gullans_PRL_2010}

In previous works~\cite{Inarrea_PRB_2007,Inarrea_APL_2007,Inarrea_APL_2009}, the interdot tunnel coupling between the quantum dots was considered incoherent, namely, the system was assumed to be in the sequential interdot tunnelling regime. However, in the present work we focus on the resonant tunnel regime and we consider coherently coupled quantum dots (\fref{dqd}). We have found in this case a different behaviour both for the current and the induced nuclear spin polarization with respect to previous works. In the coherent coupling regime, in addition to the two spin parallel triplets, there is a spin antiparallel triplet. Thus, in addition to the spin-flip transitions between the singlet and triplet states, spin-flip processes between triplet states are also present, and contribute to the nuclear spin polarization and the electronic leakage current. Furthermore, in the present model we find that the transition rates between the DQD and the leads depends on the nuclear spin polarization (\fref{sqm}). This effect will lead to new features in the current through the DQD.

The paper is organized as follows. The Hamiltonian for the DQD coupled to the leads and the surrounding nuclei spins, and the rate equations for the occupation of DQD levels and the nuclei spin polarization, are considered in~\sref{mdl}. The results and discussions are considered in~\sref{rslts}. And finally, the conclusions are presented in~\sref{clns}.

%% file: model.tex
\section{Model} \label{mdl}

\subsection{Hamiltonian and DQD eigenstates} \label{mdl_ham}

The system we investigate is a DQD coupled to two uncorrelated electron reservoirs (leads) and to the nuclei spins of the surrounding host material (\fref{dqd}). We consider a spin up and a spin down level in each dot. The Hamiltonian is the following:
\begin{eqnarray} \label{ham}
\hat{H} = \hat{H}_{DQD} + \hat{H}_{leads} + \hat{V}_{LR} + \hat{V}_T + \hat{V}_{HF}
\end{eqnarray}
where $\hat{H}_{DQD}$ and $\hat{H}_{leads}$ are the Hamiltonians for the isolated DQD and the electron reservoirs, respectively, $\hat{V}_{LR}$ is the inter-dot tunnelling Hamiltonian, and $\hat{V}_T$ and $\hat{V}_{HF}$ correspond to the DQD coupling with the leads, and the HF interaction with the nuclei spins, respectively. We neglect cotunnelling processes because we consider that the tunnelling coupling through the contact barriers is much smaller than the thermal energy and the bias voltage~\cite{Bruus}. Moreover, the energy of the DQD levels are strongly detunned respect to the contacts chemical potentials~\cite{Johnson_PRB_2005,Vorontsov_PRL_2008}. 

We consider contact HF interaction, as we regard electronic wave functions with $s$ like symmetry~\cite{Fischer_PRB_2009}. $\hat{H}_{DQD}$ and $\hat{V}_{LR}$ are:
\begin{eqnarray} \label{ham_2}
\hat{H}_{DQD} = \sum_{l \sigma} \epsilon_l \hat{n}_{l \sigma} +
\sum_l U_l \hat{n}_{l \uparrow} \hat{n}_{l \downarrow} + U_{LR}
\sum_{\sigma \sigma'} \hat{n}_{L \sigma}\hat{n}_{R \sigma'} + \sum_l g \mu_B B_{{\rm ext}} \hat{S}_{lz} \nonumber
\\ \hat{V}_{LR} = t_{LR} \sum_{\sigma} \left( \hat{d}_{L
  \sigma}^{\dagger} \hat{d}_{R \sigma} + {\rm h.c} \right) \nonumber
\end{eqnarray}
where $l = L$ (left dot), $R$ (right dot) and $\sigma = \uparrow, \downarrow$. $\hat{d}_{l \, \sigma}^{\dagger}$ ($\hat{d}_{l \, \sigma}$) creates (annihilates) an electron with spin $\sigma$ and energy $\epsilon_l$ in the $l$-th dot. $\hat{n}_{l\sigma} = \hat{d}_{l \sigma}^{\dagger} \hat{d}_{l\sigma}$ is the occupation number operator and $\hat{\mathbf{S}}_l$ the electron spin operator. $U_l$ ($U_{LR}$) is the intra-dot (inter-dot) Coulomb interaction, $B_{{\rm ext}}$ the external magnetic field and $t_{LR}$ the tunnelling matrix element between the dots. We do not focus on any particular material, thus we take $g = 2$. $\hat{H}_{leads}$ and $\hat{V}_T$ are:
\begin{eqnarray} \label{Hleads_VT}
\hat{H}_{leads} = \sum_{lk \sigma} \epsilon_{lk\sigma}
\hat{c}_{lk\sigma}^{\dagger} \hat{c}_{lk\sigma} \nonumber \\ \hat{V}_T
= \sum_{lk \sigma} \left( \gamma_{lk} \hat{c}_{lk\sigma}^{\dagger}
\hat{d}_{l \sigma} + {\rm h.c.}  \right)
\end{eqnarray}
where $\hat{c}_{lk \sigma}^{\dagger}$ ($\hat{c}_{lk \sigma}$) creates (annihilates) an electron in the $l$-th lead with momentum $\mathbf{k}$, spin $\sigma$ and energy $\epsilon_{lk\sigma}$. $\gamma_{lk}$ are the tunnelling matrix elements between the dots and the contacts. Finally, the HF interaction term $\hat{V}_{HF}$ is:
\begin{eqnarray} \label{VHF}
\hat{V}_{HF} = \sum_{l = L,R} \sum_{i = 1}^N A_i^l \, \hat{\mathbf{S}}_l \cdot\hat{\mathbf{I}}_i
\end{eqnarray}
where $A_i^l = \nu A |\Psi_0^l(\mathbf{r}_i)|^2$ is the HF coupling~\cite{Merkulov_PRB_2002} between the $i$-th nuclei spin $\hat{\mathbf{I}}_i$ at site $\mathbf{r}_i$ and the electron spin $\hat{\mathbf{S}}_l$, where $\nu$ is the volume of a unit cell containing one nuclear spin, $A$ characterizes the hyperfine coupling strength and $\Psi_0^l(\mathbf{r}_i)$ is the single-particle electronic wave function for dot $l$, evaluated at site $\mathbf{r}_i$. For simplicity, we only consider spin-$\frac12$ nuclear species. The HF couplings for the left and right dots depend on the square modulus of the electron wave functions at the position of the nuclei, therefore, for realistic dots they are in principle different for each dot~\cite{Coish_PRB_2005,Erbe_PRB_2010}. We consider the case of homogeneous hyperfine couplings~\cite{Khaetskii_PRB_2003} within each dot, hence, $A_i^l = A_l/N$, where $N$ is total number of nuclear spins, and split $\hat{V}_{HF}$ (see~\eref{VHF}) in the following two terms:
\begin{eqnarray} \label{VHF}
\hat{V}_{HF} = \begin{array}{ccc} \sum_{l = L,R} \frac{A_l}{N} \sum_{i = 1}^N \left( \frac{1}{2} \left( \hat{S}_{l+} \hat{I}_{i-} + \hat{S}_{l-} \hat{I}_{i+} \right) + \hat{S}_{lz} \hat{I}_{iz} \right) = \hat{V}_{sf} + \hat{V}_z \end{array},
\end{eqnarray}
where $\hat{S}_{\pm} = \hat{S}_x \pm \imath \hat{S}_y$ and $\hat{I}_{\pm} = \hat{I}_x \pm \imath \hat{I}_y$ are the raising and lowering operators for the electron and the nuclei spins, respectively. $\hat{V}_{sf}$ corresponds to the $x$-$y$-components, which are perpendicular to the external magnetic field and, thus, is responsible for the flip-flop transitions between electron and nuclei spins; and $\hat{V}_z$ corresponds to the $z$-component, which is parallel to the external field and, hence, contributes to the Zeeman splitting, as we will see below. We treat these two terms separately. For $\hat{V}_z$ we perform a mean-field approximation which gives~\cite{Inarrea_PRB_2007}:
\begin{eqnarray}
\hat{V}_z & \rightarrow & \begin{array}{ccc} \hat{V}_z^{MF} & = & \frac{1}{2} \sum_l A_l \hat{S}_{lz} P \end{array}
\end{eqnarray}
where $P = (\langle N_{\uparrow} \rangle - \langle N_{\downarrow} \rangle)/N$ is the net nuclear spin polarization, and $\langle N_{\uparrow} \rangle$ ($\langle N_{\downarrow} \rangle$) is the average number of spin up (down) nuclei. $\hat{V}_z^{MF}$ is now equivalent to an effective magnetic field within each dot induced by the nuclei on the electrons (the Overhauser field), and proportional to the nuclear polarization given by:
\begin{eqnarray} \label{Bnuc}
B_{\rm nuc}^{L (R)} = \frac{A_{L(R)} P}{2g \mu_B}.
\end{eqnarray}
This effective field is in general different for each dot~\cite{Jouravlev_PRL_2006}, and gives rise to an effective Zeeman splitting within each dot that adds to the one produced by the external field. Below we shall see that this inhomogeneity is essential for lifting SB~\cite{Koppens_Science_2005}. We consider $\hat{V}_T$ and $\hat{V}_{sf}$ as perturbations. Thus, the unperturbed Hamiltonian for the isolated DQD is the following:
\begin{eqnarray} \label{H0}
\hat{H}_0 = \hat{H}_{DQD} + \hat{V}_{LR} + \hat{V}_z^{MF}.
\end{eqnarray}

SB occurs when the source-drain voltage is tuned so that the number of electrons in the DQD varies between one and two. The right quantum dot is always occupied with one electron while another electron can tunnel from the source to the drain through the DQD. The transport scheme is the following:
\begin{equation} \label{SB_1}
\begin{array}{ccccc} (0,\sigma) & \longrightarrow & (\sigma',\sigma) & \longrightarrow & \left\{ \begin{array}{ccccc} (0,\uparrow \downarrow) & \longrightarrow & (0,\sigma'') & {\rm if} & \sigma \ne \sigma' \\ (\sigma,\sigma) & ({\rm SB}) & & {\rm if} & \sigma = \sigma' \end{array} \right., \end{array}
\end{equation}
where in $(n,m)$, $n$ ($m$) accounts for the population of the left (right) dot level, and $\sigma, \sigma', \sigma'' = \uparrow, \downarrow$. Interdot tunnelling ($\hat{V}_{LR}$) is allowed only between states with the same total spin. Since the state $(0,\uparrow \downarrow)$ has total spin zero, while the states with $\sigma = \sigma'$ have total spin one, the transitions $(\sigma,\sigma) \to (0,\uparrow \downarrow)$ are forbidden. The Hilbert space that we have considered, consists of the {\it atomic basis} $\{ |0,\uparrow \rangle$, $|0,\downarrow \rangle$, $|\uparrow,\downarrow \rangle$, $|\downarrow,\uparrow \rangle$, $|\uparrow,\uparrow \rangle$, $|\downarrow,\downarrow \rangle$, $|0,\uparrow \downarrow\rangle \}$. We investigate the coherent {\it resonant transport} regime. In this regime, the energies of the DQD two electrons states $|\sigma,\sigma'\rangle$ and $|0,\uparrow \downarrow\rangle$ are degenerate in the absence of a magnetic field, i.e., the so called zero detuning regime. In this case $\hat{H}_0$ (see~\eref{H0}) is exactly diagonalizable. Its eigenenergies and eigenstates are the following:
\begin{eqnarray} \label{egn_enrgs}
E_{T_x} = \epsilon_L + \epsilon_R + U_{LR} \nonumber \\ E_{S_{\pm}} = \epsilon_L + \epsilon_R
+ U_{LR} \pm \sqrt{2} \, \mathcal{N} t_{LR} \nonumber \\ E_{T_{\pm}} = \epsilon_L + \epsilon_R + U_{LR} \pm \left( g\mu_B B_{{\rm ext}} + \frac{A_+}{2} \, P \right)
\end{eqnarray}
and:
\begin{eqnarray} \label{egn_vctrs}
| T_+ \rangle = | \uparrow, \uparrow \rangle \nonumber \\
| T_- \rangle = | \downarrow, \downarrow \rangle \nonumber \\
| T_x \rangle = \frac{1}{\mathcal{N}} \left( | T_0 \rangle - x |S_{02} \rangle \right) \nonumber \\ | S_{\pm} \rangle = \frac{1}{\sqrt{2}} \left( | S_{11} \rangle \pm \frac{1}{\mathcal{N}} \left( | S_{02} \rangle + x |T_0 \rangle \right) \right)
\end{eqnarray}
where:
\begin{eqnarray}
| T_0 \rangle = \frac{1}{\sqrt{2}} \, (| \uparrow, \downarrow \rangle + | \downarrow, \uparrow \rangle) \nonumber \\ | S_{11} \rangle = \frac{1}{\sqrt{2}} \, (| \uparrow, \downarrow \rangle - | \downarrow, \uparrow \rangle ) \nonumber \\ | S_{02} \rangle = | 0, \uparrow \downarrow \rangle
\end{eqnarray}
being:
\begin{eqnarray} \label{x}
\mathcal{N} = \sqrt{1 + x^2} \nonumber \\ x = \frac{1}{\sqrt{2}} \, \frac{A_- P}{2 \, t_{LR}} \nonumber \\ A_{\pm} = \frac{1}{2} \, (A_L \pm A_R).
\end{eqnarray}
Thus, the two electron {\it molecular basis} is $\{ |\pm \rangle, |S_{\pm}\rangle, |T_x\rangle, | T_{\pm} \rangle \}$, where $|+(-) \rangle = | 0, \uparrow(\downarrow) \rangle$ are the single electron states. \Eref{egn_vctrs} shows that $\hat{V}_{LR}$ mixes $S_{11}$ and $S_{02}$ singlets, and $\hat{V}_z^{MF}$ mixes $T_0$ triplet with $S_{11}$ and $S_{02}$ singlets when $A_L \ne A_R$. The singlet-triplet (ST) mixing is given by the weight $x/\mathcal{N}$. This quantity is the ratio between the Zeeman splitting difference within each dot ($A_-P/2$), and the {\it exchange energy} defined as $|E_{S_{\pm}} - E_{T_x}| = \sqrt{2} \, \mathcal{N} \, t_{LR}$. Therefore, the ST mixing depends on the competition between these two energy scales. A large (small) difference between the HF coupling intensities, and small (large) interdot tunnelling strength increases (decreases) the ST mixing (see~\eref{egn_vctrs} and~\eref{x}). Furthermore, $x$ is zero when either the HF couplings have the same value for both dots ($A_L = A_R$) or the nuclei spins are completely depolarized ($P = 0$). In both cases the usual singlet-triplet basis is recovered.  Due to the mixing with $T_0$ triplet state, $S_{\pm}$ are not pure singlet states anymore, however, for simplicity we shall continue calling them {\it singlet} states. Finally, notice that now the eigenenergies of singlet states also depend on the nuclear spin polarization through $\mathcal{N}$ (see~\eref{egn_enrgs} and~\eref{x}).

The transport scheme in the molecular basis is the following:
\begin{eqnarray} \label{SB_2}
(0,\sigma) & \longrightarrow & \left\{ \begin{array}{lcl} S_{\pm} & \longrightarrow & (0,\sigma') \\ T_x & \longrightarrow & \left\{ \begin{array}{ccc} (0,\sigma') & {\rm if} & x \ne 0 \\ T_0 \, ({\rm SB}) & {\rm if} & x = 0 \end{array} \right. \\ T_{\pm} & ({\rm SB}) & \end{array} \right.
\end{eqnarray}
This transport scheme shows that for the coherent interdot tunnelling regime, when having different Overhauser fields in the dots, a current channel ($(0,\sigma) \to T_x \to (0,\sigma')$) is opened. Therefore, there are two possible situations: i) $x = 0$. In this case the incoming electron will fall either in a singlet state ($S_{\pm}$) or in a triplet state ($T_0$ or $T_{\pm}$). Thus, there are {\it two} transport channels and {\it three} SB states; ii) $x \ne 0$. The incoming electron will fall either in a singlet state ($S_{\pm}$), the $T_x$ state or in the $T_{\pm}$ triplets. Therefore, now there are {\it three} transport channels and {\it two} SB states. However, once the electrons drop in a SB state, the current drops to zero.

\subsection{Rate equations} \label{mdl_rteqs}

The Hamiltonian~\eref{ham} can be written now as follows:
\begin{eqnarray}
\hat{H} = \hat{H}_{0} + \hat{V}_T + \hat{V}_{sf} + \hat{H}_{leads},
\end{eqnarray}
where the eigenstates of $\hat{H}_0$ (see~\eref{egn_vctrs}) are the unperturbed states. $\hat{V}_T$ induces transitions between the leads and the DQD, namely, between one-electron and two-electron states. $\hat{V}_{sf}$ is responsible for the spin flip-flop transitions between the DQD electron spins and the surrounding nuclei spins. The time evolution of the DQD molecular states are obtained with the following rate equations: 
\begin{eqnarray} \label{rt_eqns}
\fl \dot{\rho}_{T_{\pm}} = W_{T_{\pm},S_+} \rho_{S_+} + W_{T_{\pm},S_-} \rho_{S_-} + W_{T_{\pm},T_x} \rho_{T_x} + \Gamma_{T_{\pm},\pm} \rho_{\pm} \nonumber \\ - (W_{S_+,T_{\pm}} + W_{S_-,T_{\pm}} + W_{T_x,T_{\pm}}) \rho_{T_{\pm}} \nonumber \\ \fl \dot{\rho}_{T_x} = W_{T_x,T_+} \rho_{T_+} + W_{T_x,T_-} \rho_{T_-} + \Gamma_{T_x,+} \rho_{+} + \Gamma_{T_x,-} \rho_{-} \nonumber \\ - (W_{T_+,T_x} + W_{T_-,T_x} + \Gamma_{+,T_x} + \Gamma_{-,T_x}) \rho_{T_x} \nonumber \\ \fl \dot{\rho}_{S_{\pm}} = W_{S_{\pm},T_+} \rho_{T_+} + W_{S_{\pm},T_-} \rho_{T_-} + \Gamma_{S_{\pm},+} \rho_{+} + \Gamma_{S_{\pm},-} \rho_{-} \nonumber \\ - (W_{T_+,S_{\pm}} + W_{T_-,S_{\pm}} + \Gamma_{+,S_{\pm}} + \Gamma_{-,S_{\pm}}) \rho_{S_{\pm}} \nonumber \\ \fl \dot{\rho}_{\pm} = \Gamma_{\pm,S_+} \rho_{S_+} + \Gamma_{\pm,S_-} \rho_{S_-} + \Gamma_{\pm,T_x} \rho_{T_x} - (\Gamma_{S_+,\pm} + \Gamma_{S_-,\pm} + \Gamma_{T_x,\pm} + \Gamma_{T_{\pm},\pm}) \rho_{\pm}
\end{eqnarray}
where $\rho_i$ is the occupation of the $i$-th state. $\Gamma_{i,f}$ ($W_{i,f}$) are the tunnelling (spin-flip) rates between an initial DQD state $|i\rangle$ and a final DQD molecular state $|f\rangle$. Both tunnelling and spin-flip rates are computed in~\sref{mdl_rts}. 

The spin electron-nuclei flip-flop processes induce a non-negligible nuclear spin polarization, which will be positive or negative depending on the specific spin-flip processes. In the following scheme we show the spin-flip processes which contribute to each nuclear spin polarization direction:
\begin{eqnarray} \label{P_sign}
T_+ \rightarrow \{S_+,S_-,T_x\} & \Rightarrow & \dot{P} > 0 \nonumber \\
T_- \rightarrow \{S_+,S_-,T_x\} & \Rightarrow & \dot{P} < 0 \nonumber \\
\{S_+,S_-,T_x\} \rightarrow T_+ & \Rightarrow & \dot{P} < 0 \nonumber \\
\{S_+,S_-,T_x\} \rightarrow T_- & \Rightarrow & \dot{P} > 0. 
\end{eqnarray}
E.g., the process $T_+ \to \{S_+,S_-,T_x\}$ flips down an electron spin and up a nuclear spin, thus, polarizes positively the nuclei spins. Furthermore, the nuclear spin polarization becomes dynamical due to the electrons tunnelling through the DQD. Therefore, we describe the time evolution for the induced nuclear spin polarization using the following rate equation:
\begin{eqnarray} \label{P_eqn}
\fl \dot{P} = \left( W_{T_-,S_+} - W_{T_+,S_+} \right) \rho_{S_+} + \left( W_{T_-,S_-} - W_{T_+,S_-} \right) \rho_{S_-} + \left( W_{T_-,T_x} - W_{T_+,T_x} \right) \rho_{T_x} \nonumber \\ + (W_{S_+,T_+} + W_{S_-,T_+} + W_{T_x,T_+}) \rho_{T_+} - (W_{S_+,T_-} + W_{S_-,T_-} + W_{T_x,T_-}) \rho_{T_-} \nonumber \\ - W_{{\rm rel}} \, P
\end{eqnarray}
where $W_{{\rm rel}}$ is a phenomenological rate that accounts for the nuclear dipole-dipole spin interaction which is responsible for nuclear spin depolarization. In this equation we have assumed that in absence of spin-flip processes ($W_{f,i}$ all zero), the nuclei spins completely depolarize, namely, that temperature is much larger than the nuclei spin level splittings. As will be shown in~\sref{mdl_rts}, both the tunnelling and the spin-flip rates depend on the nuclear spin polarization. Therefore,~\eref{rt_eqns} and~\eref{P_eqn} form a set of eight non-linear equations which we must be solved numerically (\sref{rslts}).

\subsubsection{Reduced model} \label{eff_mdl}

In order to get physical insight on the results which will be discussed in \sref{rslts}, we have developed a simplified model for the rate equations~\eref{rt_eqns} and~\eref{P_eqn} through the following assumptions: i) we consider that spin-flip processes are effective when the electrons are in a SB state. Thus, once the electrons are in a singlet state it is much more probable for them to tunnel out of the DQD than to flip its spin to a $T_{\pm}$ triplet state. Therefore, we neglect the singlet to triplet spin-flip transitions ($W_{T_{\pm},S_+}$ and $W_{T_{\pm},S_-}$). ii) The singlet states and the one electron states empty much faster than the $T_{\pm}$ triplets (SB) and the $T_x$ state. Thus, in the time scale on which spin-flip transitions are relevant we assume that the occupation of the one electron state and of the singlet states have reached their stationary value, namely, $\dot{\rho}_{\pm} \approx \dot{\rho}_{S_{\pm}} \approx 0$. iii) The nuclei relaxation time due to the spin dipole-dipole interaction is long enough to be neglected ($W_{\rm rel} \to 0$). Under these conditions, the rate equation for the nuclei spin polarization~\eref{P_eqn} is related to the rate equations for the triplet states~\eref{rt_eqns} through the relation $\dot{\rho}_{T_-} - \dot{\rho}_{T_+} = \dot{P}$, thus, $\rho_{T_-} - \rho_{T_+} = P$ (where we consider $\rho_{T_-}(0) - \rho_{T_+}(0) - P(0) = 0$ and $\rho_{\pm}(0) = \rho_{S_{\pm}}(0) = 0$ as initial condition). Taking these considerations into account the rate equations~\eref{rt_eqns} become:
\begin{subequations}
\begin{eqnarray}
\dot{\rho}_T = \left( 2\omega_+^{tt} + \beta \right) \rho_{T_x} - (\omega_+^{st}  + \omega_+^{tt}) \rho_T + (\omega_-^{st}  + \omega_-^{tt}) P \label{rhoT} \\ 
\dot{\rho}_{T_x} = (\omega_+^{st}  + \omega_+^{tt}) \rho_T - (\omega_-^{st}  + \omega_-^{tt})_- P - (2\omega_+^{tt} + \beta ) \rho_{T_x} \label{rhoTx} \\
\dot{P} = 2\omega_-^{tt} \rho_{T_x} + (\omega_-^{st}  + \omega_-^{tt}) \rho_T - (\omega_+^{st}  + \omega_+^{tt}) P
\end{eqnarray}
\end{subequations}
where $\rho_T = \rho_{T_+} + \rho_{T_-}$ and,
\begin{eqnarray} \label{abW}
\omega_{\pm}^{tt} = \frac{1}{2} \left( W_{T_-,T_x} \pm W_{T_+,T_x} \right) \nonumber \\
\omega_{\pm}^{st} = \frac{1}{2} \frac{(W_{S_+,T_+} + W_{S_-,T_+}) \pm (W_{S_+,T_-} + W_{S_-,T_-})}{1 + \Gamma/\Gamma_{+,S_+}} \nonumber \\
\beta = \frac{2 \Gamma_{+,T_x}}{1 + \Gamma_{+,S_+}/ \Gamma}.
\end{eqnarray}
We have considered the same value of the tunnelling coupling for both contact barriers, $\Gamma_L = \Gamma_R = \Gamma$ (see~\sref{tnl_rts}). Finally, by summing \eref{rhoT} and \eref{rhoTx} we find that $\rho_T + \rho_{T_x} = 1$, where we consider as initial condition: $\rho_T(0) + \rho_{T_x}(0) = 1$, thus, the rate equations become:
\begin{eqnarray} \label{rdcd_eqns}
\dot{\rho}_{T_x} = (\omega_+^{tt} + \omega_+^{st}) - (3\omega_+^{tt} + \beta + \omega_+^{st}) \rho_{T_x} - (\omega_-^{tt} + \omega_-^{st}) P \nonumber \\ \dot{P} = (\omega_-^{tt} + \omega_-^{st}) + (\omega_-^{tt} - \omega_-^{st}) \rho_{T_x} - (\omega_+^{tt} + \omega_+^{st}) P
\end{eqnarray}
The rates $\omega_{\pm}^{st}$ account for the SB lifting due to ST spin relaxation: $T_{\pm} \overset{\omega_{\pm}^{st}}{\longrightarrow} \{S_+,S_-\}$. $\omega_{\pm}^{tt}$ ($\beta$) account for the spin-flip (tunnelling) rates from $T_x$ to $T_{\pm}$ ($T_x$ to $|\pm\rangle$). In~\sref{tnl_rts} we will see that the rate $\Gamma_{+,S_+} \simeq \Gamma/2$, hence, it can be regarded as constant. In~\sref{rslts} it will be shown that the current through the DQD depends on ratio between $\omega_+^{tt}$ and $\beta$. The following scheme offers a qualitative description of this dependence for the case of $A_- \ne 0$:
\begin{eqnarray} \label{alf_bt}
T_x \longrightarrow & \left\{ \begin{array}{c} \overset{\omega_+^{tt} \ll \beta}{\longrightarrow} \\ \overset{\omega_+^{tt} \gg \beta}{\longrightarrow} \end{array} \right. & \begin{array}{ccc} (0,\sigma) & \longrightarrow & {\rm High \, \, Current \, \, Regime} \\ T_{\pm} & \longrightarrow & {\rm Low \, \, Current \, \, Regime}. \end{array}
\end{eqnarray}
When $\omega_+^{tt} \gg \beta$ electrons in $T_x$ state have a larger probability to spin-flip to the $T_{\pm}$ triplets than to tunnel out of the DQD, and only the leakage current coming from the $T_{\pm} \to \{S_+,S_-\}$ transitions provide a leakage current, which we define as the {\it low current regime}. On the contrary, when $\omega_+^{tt} \ll \beta$ electrons in $T_x$ state have a larger probability to tunnel out of the DQD than to spin-flip to the blocked $T_{\pm}$ triplets, and the current is enhanced. We define this case as the {\it high current regime}. In~\sref{rslts}, we will see that the competition between spin-flip and tunnelling transition rates determine the behaviour of the leakage current. Moreover, we will see that the transition between the low current regime and the high current regime is obtained by varying either the interdot tunnelling strength or the HF coupling intensity.

\subsection{Transition rates} \label{mdl_rts}

\subsubsection{Spin-flip rates} \label{sf_rts}

The transition rates between $\hat{H}_0$ eigenstates (see~\eref{egn_vctrs}) due to spin-flip processes are calculated in perturbation theory by means of Fermi's Golden Rule~\cite{Rudner_PRL_2007,Inarrea_PRB_2007,Bruus}. The left (right) dot electron spin $z$-projection is given by $m_{L(R)} = \pm 1/2$, hence, the total spin projection in the $z$-direction for a DQD two electron state is $M = m_L + m_R$ ($M = -1,0,1$ for triplets, and $M = 0$ for singlets). The spin-flip interaction term $\hat{V}_{sf}$ (see~\eref{VHF}) increases an electron spin by one while decreasing a nuclei spin by one also (and vice versa), thus, $M: \, -1 \overset{\hat{V}_{sf}}{\longleftrightarrow} 0 \overset{\hat{V}_{sf}}{\longleftrightarrow} 1$. We consider the initial state $|i_N\rangle|\alpha_M\rangle$ which consists of the initial nuclei states $|i_N \rangle$ and the DQD electron states $|\alpha_M \rangle$. $|i_N \rangle$ is given by $| i_N \rangle = |m_1, m_2, \ldots, m_j, \cdots, m_N \rangle$, where $m_j = \pm 1/2$ is the spin of the $j$-th nuclei, and $|\alpha_M \rangle \in \{|S_{\pm}\rangle, |T_x\rangle, | T_{\pm} \rangle \}$. The final state $|f_N\rangle|\beta_{M'}\rangle$ is connected to the initial one by having the $j$-th nuclear spin flipped and a different electronic state $|\beta_{M'} \rangle$ with $|M - M'|= 1$. Therefore the spin-flip rate for the transition that flips up an electron spin and down a nuclei spin is:
\begin{eqnarray}
\fl W_{\beta_{M+1} \alpha_M} = 2 \pi \sum_{j = 1}^N \sum_{i_N} |\langle \beta_{M+1} | \langle f_N | \hat{V}_{sf} | i_N \rangle | \alpha_M \rangle |^2 \, \mathcal{W}_{i_N} \, \delta \left( E_{\beta_{M+1}} - \delta_j - E_{\alpha_M} \right)  \\ \fl = 2 \pi \frac{1}{N^2} \frac{1}{2^2} \left| \sum_{l = L,R} A_l \langle \beta_{M+1} | \hat{S}_l^+ | \alpha_M \rangle \right|^2 \sum_{j = 1}^N \sum_{i_N} \langle i_N | \hat{I}_j^+ \hat{I}_j^- | i_N \rangle \, \mathcal{W}_{i_N} \, \delta \left( E_{\beta_{M+1}} - \delta_j - E_{\alpha_M} \right). \nonumber
\end{eqnarray}
where $|f_N\rangle = \hat{I}_j^- |i_N\rangle$ and we have taken $\hbar = 1$. The sum over initial states runs over all configurations of the internal degrees of freedom, $i_N$, that give the state $| i_{N} \rangle$. Each state is weighted by the probability of having that configuration, which is given by the distribution function $\mathcal{W}_{i_N}$. $E_{\alpha_M}$ ($E_{\beta_{M + 1}}$) is the energy of the initial (final) electronic state $|\alpha_M \rangle$ ($|\beta_{M + 1} \rangle$), and $\delta_j$ is the energy splitting between the up-down levels of the $j$-th nuclei spin. We assume independent nuclei spins, hence, $\mathcal{W}_{i_N} = \mathcal{W}_{m_1} \times \mathcal{W}_{m_2} \times \ldots \times \mathcal{W}_{m_N}$, where $\mathcal{W}_{m_j}$ is the probability that nuclear spin $j$ has the value $m_j$. Thus, the sum over initial nuclear states becomes:
\begin{eqnarray}
\sum_{i_N} \langle i_N | \hat{I}_j^+ \hat{I}_j^- | i_N \rangle \, \mathcal{W}_{i_N} = \sum_{m_j} \langle m_j | \hat{I}_j^+ \hat{I}_j^- | m_j \rangle \, \mathcal{W}_{m_j}
\end{eqnarray}
where we have used the normalization condition $\mathcal{W}_{m_k = 1/2} + \mathcal{W}_{m_k = -1/2} = 1$ to eliminate all other sums over $m_k$ except $m_j$. The probability of having nuclear spin $j$ in a certain state is related to the overall nuclear spin polarization as:
\begin{eqnarray} \label{Wm}
\mathcal{W}_{m_j = 1/2} = \frac{\langle N_{\uparrow} \rangle}{N} = \frac{1 + P}{2} \nonumber \\ \mathcal{W}_{m_j = -1/2} = \frac{\langle N_{\downarrow} \rangle}{N} =
\frac{1 - P}{2}.
\end{eqnarray}
In general, the $g$-factor is much smaller for nuclei spins than for electrons~\cite{Slichter}. Therefore, under experimental conditions~\cite{Ono_Science_2002,Pfund_PRL_2007,Shaji_Nature_2008,Churchill_Nature_2009,Koppens_PRL_2007}, the nuclear splitting is usually negligible compared to temperature and the energy difference between electronic levels, hence, $\delta_j$ $\forall \, j = 1 \ldots N$ can be safely neglected. Thus, the spin-flip rate becomes:
\begin{eqnarray}
\fl W_{\beta_{M+1} \alpha_M} = \frac{\pi}{2N} \left| \sum_{l = L,R} A_l \langle \beta_{M+1} | \hat{S}_l^+ | \alpha_M \rangle
\right|^2 \frac{1 + P}{2} \, \delta \left( E_{\beta_{M+1}} - E_{\alpha_M} \right).
\end{eqnarray}
Repeating the same procedure we have that the spin-flip rate for the transition that flips down an electron spin and up a nuclei spin is the following:
\begin{eqnarray}
\fl W_{\beta_{M-1} \alpha_M} = \frac{\pi}{2N} \left| \sum_{l = L,R} A_l \langle \beta_{M-1} | \hat{S}_l^- | \alpha_M \rangle
\right|^2 \frac{1 - P}{2} \, \delta \left( E_{\beta_{M-1}} - E_{\alpha_M} \right).
\end{eqnarray}
Notice that these spin flip-rates depend on how much the nuclei are polarized. When the nuclei spins are fully polarized in the positive (negative) direction $W_{\beta_{M-1} \alpha_M}$ ($W_{\beta_{M+1} \alpha_M}$) vanishes.

We see that the derived spin-flip rate requires energy conservation, so strictly speaking leads to zero spin-flip for $E_{\alpha_{M}} \ne E_{\beta_{M \pm 1}}$. However, in reality it is possible to exchange energy~\cite{Fujisawa_Science_1998} with the environment e.g. as phonons. We model this by replacing the Dirac delta by the following expression:
\begin{eqnarray}
\delta \left( E_{\beta_{M \pm 1}} - E_{\alpha_M} \right) & \rightarrow & \frac{1}{\pi} \frac{\gamma}{\left( E_{\beta_{M \pm 1}} - E_{\alpha_M} \right)^2 + \gamma^2} \times \mathcal{C}_{\beta_{M \pm 1} \alpha_M}
\end{eqnarray}
where:
\begin{eqnarray} \label{C}
\fl \mathcal{C}_{\beta_{M \pm 1} \alpha_M} = \left\{ \begin{array}{ccccl} 1 & {\rm if} & E_{\beta_{M \pm 1}} > E_{\alpha_M} & \rightarrow & {\rm Energy \, \,  emission} \\ \exp \left( \frac{E_{\beta_{M \pm 1}} - E_{\alpha_M}}{k_B T} \right) & {\rm if} & E_{\beta_{M \pm 1}} < E_{\alpha_M} & \rightarrow & {\rm Energy \, \, absorption} \end{array} \right.
\end{eqnarray}
where $T$ is the temperature and $k_B$ the Boltzmann constant. The Lorentzian is maximal for the elastic case and falls off with increasing energy exchange on the characteristic scale $\gamma$. This parameter is assumed to be of the order of the typical phonon energy, $\gamma \sim \, \mu$eV~\cite{Fujisawa_Science_1998}. The function $\mathcal{C}_{\beta_{M \pm 1} \alpha_M}$ accounts for the low temperature energy emission/absorption asymmetry~\cite{Fujisawa_Science_1998,Blum}, i.e., it is much easier to emit than absorb energy from e.g. a phonon bath. Formally, one can include the electron-phonon coupling as a perturbation together with $\hat{V}_{sf}$ and thereafter use a $T$-matrix approach to find the phonon-mediated HF spin-flip rate~\cite{Kim_PRB_1994,Erlingsson_PRB_2001,Prada_PRB_2008}. However, here we do not pursue an exact modeling of the way the energy is exchanged with the environment, but simply include the fact that the spin-flip rate decreases as the energy involved in the flip-flop processes increases~\cite{Rudner_PRL_2007,Inarrea_PRB_2007}. Therefore, the spin-flip rates we obtain are the following:
\begin{eqnarray} \label{Wsf}
\fl W_{\beta_{M \pm 1} \alpha_M} = \frac{1}{2N} \left| \sum_{l = L,R} A_l \langle \beta_{M \pm 1} | \hat{S}_l^{\pm} | \alpha_M \rangle \right|^2 \frac{1 \pm P}{2} \, \frac{\gamma}{\left( E_{\beta_{M \pm 1}} - E_{\alpha_M} \right)^2 + \gamma^2} \, \mathcal{C}_{\beta_{M \pm 1} \alpha_M}.
\end{eqnarray}
In~\sref{rslts}, we will show that the amount of induced nuclear spin polarization depends on the competition between emission and absorption processes. The energies that appear in the function $\mathcal{C}_{\beta_{M \pm 1} \alpha_M}$ (see~\eref{C}) are the eigenenergies shown in~\eref{egn_enrgs}. Thus, the energy differences in $\mathcal{C}_{\beta_{M \pm 1} \alpha_M}$ depend on the interdot tunnel strength, the HF coupling and the external magnetic field. Therefore, for a given temperature the emission/absorption asymmetry can be controlled through these parameters. Finally, the only matrix elements between the DQD states which are different from zero in~\eref{Wsf} are:
\begin{subequations}
\begin{eqnarray}
\langle T_x | \left( A_L \hat{S}_L^{\pm} + A_R \hat{S}_R^{\pm} \right) | T_{\mp} \rangle = \frac{\sqrt{2} \, A_+}{\mathcal{N}} \label{TT_rel} \\
\langle S_{\pm} | \left( A_L \hat{S}_L^- + A_R \hat{S}_R^- \right) | T_+ \rangle = -\left( A_- \mp \frac{x A_+}{\mathcal{N}} \right) \label{STa_rel} \\
\langle S_{\pm} | \left( A_L \hat{S}_L^+ + A_R \hat{S}_R^+ \right) | T_- \rangle = A_- \pm \frac{x A_+}{\mathcal{N}} 
\label{STb_rel}
\end{eqnarray}
\end{subequations}
From these expressions we distinguish between two different HF mediated spin relaxation processes: i) the triplet-triplet relaxation~\eref{TT_rel}, and ii) the singlet-triplet relaxation (\eref{STa_rel} and~\eref{STb_rel}). Notice that if the HF coupling intensities have the same value for each dot ($A_- = 0$, and then $x = 0$) only the triplet-triplet spin relaxation survives ($T_{\mp} \to T_x = T_0$) while ST relaxation probabilities become zero. Thus, SB lifting involves having $A_- \ne 0$, as discussed in~\sref{mdl_ham} (see~\eref{SB_2}). Finally, the matrix elements also depend on the interdot tunnel through the ST mixing parameter $x$ (see~\eref{x}).

\subsubsection{Tunnelling rates} \label{tnl_rts}

The tunnelling rates between the leads and the DQD are calculated using Fermi's Golden Rule~\cite{Bruus,Blum}. Given an initial DQD state with $n$ electrons $| \alpha_n \rangle$ the tunnelling rate for an incoming electron to the final DQD state $|\beta_{n + 1} \rangle$ with $n + 1$ electrons is:
\begin{eqnarray}
\Gamma_{\beta_{n + 1},\alpha_n}^l = \Gamma_l f(\mu_D - \mu_l) \sum_{\sigma} | \langle \beta_{n + 1} | \hat{d}_{l\sigma}^{\dagger} | \alpha_n \rangle |^2.
\end{eqnarray}
And the tunnelling rate for an outcoming electron to the final DQD state $|\beta_{n - 1} \rangle$  with $n - 1$ electrons is:
\begin{eqnarray}
\Gamma_{\beta_{n - 1},\alpha_n}^l = \Gamma_l \left( 1 - f(\mu_D - \mu_l) \right) \sum_{\sigma} | \langle \beta_{n - 1} | \hat{d}_{l\sigma} | \alpha_n \rangle |^2
\end{eqnarray}
where $l = L,R$. $\Gamma_l = 2 \pi \, |\gamma_{lk}|^2 \, D_l$, where it is assumed that the density of states in both leads $D_l$ and the the tunnelling couplings $\gamma_{lk}$ (see~\eref{Hleads_VT}) are energy-independent. $f(\mu_D - \mu_{L(R)})$ is the Fermi distribution function for the left (right) lead, $\mu_{L(R)}$ is the chemical potential of the left (right) lead and $\mu_D$ is the DQD chemical potential. The DQD states that appear in the matrix elements in the tunnelling rates are the eigenstates of $\hat{H}_0$ (see~\eref{egn_vctrs}). The chemical potentials ($\mu_l$) are tuned such that the system is in the SB regime, and we define $\Gamma_{\beta_{n \pm 1},\alpha_n} = \sum_{l = L,R} \Gamma_{\beta_{n \pm 1},\alpha_n}^l$. Therefore, the tunnelling rates different from zero are:
\begin{eqnarray}
\Gamma_{T_{\pm},\pm} = \Gamma_L \nonumber \\
\Gamma_{S_+,\pm} = \frac{\Gamma_L}{4} \left( 1 + \frac{x^2}{\mathcal{N}^2} \mp \frac{2x}{\mathcal{N}} \right) \nonumber \\
\Gamma_{S_-,\pm} = \frac{\Gamma_L}{4} \left( 1 + \frac{x^2}{\mathcal{N}^2} \pm \frac{2x}{\mathcal{N}} \right) \nonumber \\
\Gamma_{T_x,\pm} = \frac{\Gamma_L}{2} \frac{1}{\mathcal{N}^2}.
\end{eqnarray}
for electrons tunnelling from the leads into the DQD, and:
\begin{eqnarray} \label{rrts}
\Gamma_{\pm,S_+} = \Gamma_{\pm,S_-} = \frac{\Gamma_R}{2} \frac{1}{\mathcal{N}^2} \nonumber \\
\Gamma_{\pm,T_x} = \Gamma_R \frac{x^2}{\mathcal{N}^2}
\end{eqnarray}
for electrons tunnelling out of the DQD to the leads. As pointed out in~\sref{mdl_rteqs}, the tunnelling rates also depend on the nuclear spin polarization. Additionally, $\Gamma_{\pm,T_x}$ becomes zero if $x = 0$ (see~\eref{x}), i.e., when the HF couplings within each dot are the same ($A_- = 0$), or the nuclei spins are completely depolarized ($P = 0$).

Finally, the tunnelling current through the DQD is given by~\cite{Bruus}:
\begin{eqnarray} \label{crnt}
I_l = (-e) \sum_n \sum_{\alpha_n,\beta_{n \pm 1}} \left( \Gamma_{\beta_{n+1},\alpha_n}^l - \Gamma_{\beta_{n-1},\alpha_n}^l \right) \rho_{\alpha_n}
\end{eqnarray}
where $l = L,R$.

%% file: results.tex
\section{Results and discussion} \label{rslts}

In this section, we present the results obtained by solving numerically the full system of rate equations derived in~\sref{mdl} (\eref{rt_eqns} and~\eref{P_eqn}). The system has three main relaxation time scales: i) the tunnelling through the contact barriers, ii) the electron-nuclei spin-flip, and iii) the nuclear spin relaxation ($\propto W_{\rm rel}$). The nuclear relaxation time is the largest one ($\sim$ minutes~\cite{Koppens_Science_2005,Churchill_Nature_2009,Huttel_PRB_2004}). We are mainly interested in investigating the system on a time scale for which the tunnelling and the spin-flip are the relevant relaxation times, whereas the nuclear spin relaxation time is much larger.

We will discuss the behaviour of the electronic current and the induced nuclei spin polarization versus the external magnetic field for different HF couplings and interdot tunnelling strengths. The HF interaction constant is not well known for all materials, thus, we have considered values for the HF coupling intensity in the range $A_L = 70-90 \, \mu$eV~\cite{Pfund_PRL_2007,Churchill_Nature_2009,Merkulov_PRB_2002}. The difference between the HF couplings in each dot is held constant for all cases ($A_R = 0.8A_L$). The plots described below have been obtained by solving numerically the rate equations derived in~\sref{mdl}, \eref{rt_eqns} and~\eref{P_eqn}, sweeping the external magnetic field from negative to positive values. We have performed the calculations for three cases: i) In order to observe only the effect of the dynamical nuclear spin polarization, we consider that the nuclei spins are initially fully depolarized for each value of the external magnetic field. Thus, the sweeping rate is much slower than the nuclear spin relaxation rate $W_{\rm rel}$ (see~\eref{P_eqn}). Nevertheless, for each value of the external field the rate equations are solved in a time scale much smaller than the nuclear spin relaxation rate in order to capture only the tunnelling and spin-flip events. ii) Initially the nuclei spins are completely depolarized, however, while the magnetic field is swept, nuclear spin polarization is built up. Thus, the sweeping rate is much faster than the nuclear spin relaxation rate, and a {\it feedback} processes between nuclear spins and electronic current occurs. iii) we proceed as in ii) sweeping the external field forwards and backwards. In this case we observe hysteresis, as has been widely observed experimentally for different DQD devices.~\cite{Koppens_Science_2005,Pfund_PRL_2007,Churchill_Nature_2009,Ono_PRL_2004}

\begin{figure}
  \centering 
	\subfloat[]{\label{E_rgm_a}\includegraphics[width=0.075\textwidth]{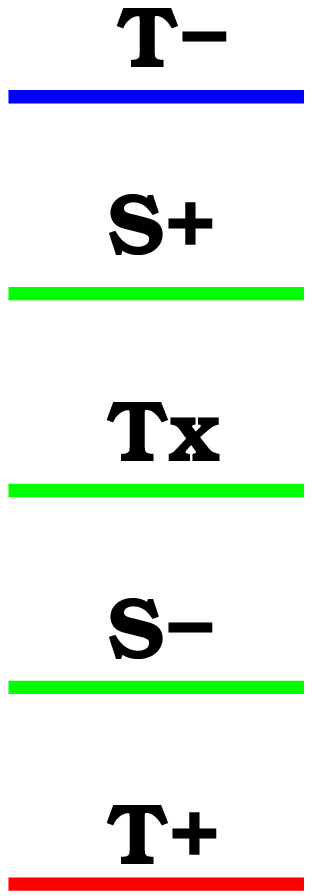}}
	\subfloat[]{\label{E_rgm_b}\includegraphics[width=0.075\textwidth]{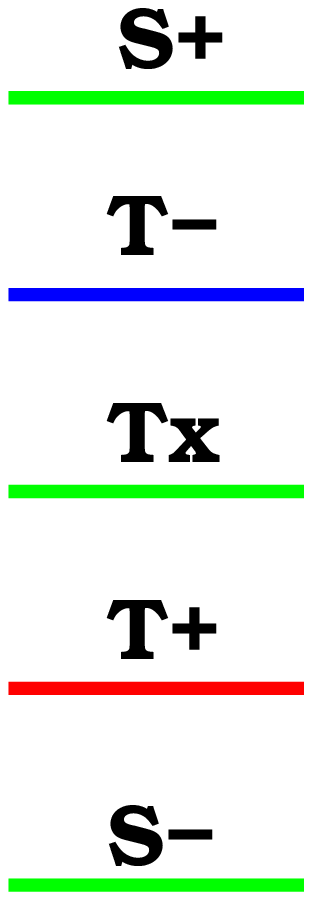}}
	\subfloat[]{\label{E_rgm_c}\includegraphics[width=0.075\textwidth]{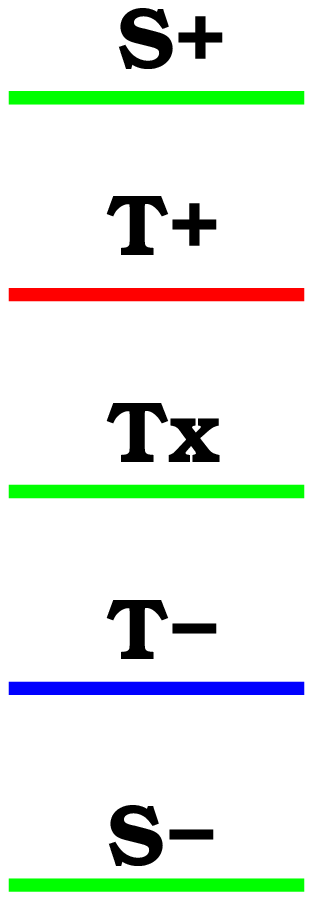}}
	\subfloat[]{\label{E_rgm_d}\includegraphics[width=0.075\textwidth]{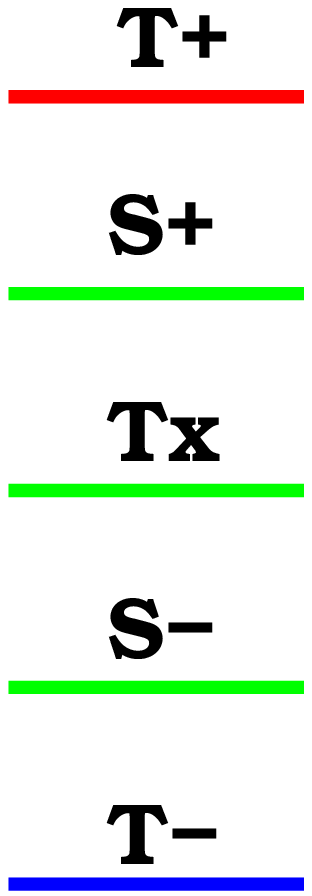}}
  \caption{(Colour online.) Schematic energy level arrangement for the DQD eigenstates in the different external magnetic field regions that are investigated. a) $B_{\rm ext} < B_{S_{\pm},T_{\mp}}$, b) $B_{S_{\pm},T_{\mp}} < B_{\rm ext} < B_{TT}$, c) $B_{TT} < B_{\rm ext} < B_{S_{\pm},T_{\pm}}$ and d) $B_{S_{\pm},T_{\pm}} < B_{\rm ext}$.}.
  \label{E_rgms}
\end{figure}

As stated in~\sref{mdl}, we consider the zero detuning case (i.e., $| \sigma, \sigma' \rangle$ and $|S_{02} \rangle$ are degenerate). In previous works~\cite{Inarrea_APL_2007,Inarrea_APL_2009}, finite detuning was considered, therefore only transitions close to one ST level crossing are important. Now, different energy levels approaches participate and we take into account all HF mediated electron spin relaxation processes, at a fixed external magnetic field. \Fref{E_rgms} shows the energy levels for the different regimes considered. $B_{S_{\pm},T_{\mp}}$ ($B_{S_{\pm},T_{\pm}}$) corresponds to the value of the external field for which the $S_{\pm}T_{\mp}$ ($S_{\pm}T_{\pm}$) crossings occur, and $B_{TT}$ the value for which the $T_{\pm}T_x$ crossing occurs. Notice that the different $S_{\pm}T_{\mp}$ crossings occur for the same value of the external field (see~\eref{egn_enrgs}), and the same happens for the different $S_{\pm}T_{\pm}$ crossings. The energy emission spin-flip rates which contribute for each range of the external field in~\fref{E_rgms} are given in the following scheme: 
\begin{subequations}
\begin{eqnarray} \label{B_rgms}
B_{\rm ext} < B_{S_{\pm},T_{\mp}} & \Rightarrow & \begin{array}{ccc} W_{S_{\pm},T_-} , W_{T_x,T_-} , W_{T_+,T_x} & \Rightarrow & P < 0 \end{array} \label{B_rgms_a} \\
B_{S_{\pm},T_{\mp}} < B_{\rm ext} < B_{TT} & \Rightarrow & \left\{ \begin{array}{rcc} W_{S_-,T_-} , W_{T_x,T_-} , W_{T_+,T_x} & \Rightarrow & P < 0 \\ W_{S_-,T_+} & \Rightarrow & P > 0 \end{array} \right. \label{B_rgms_b} \\
B_{TT} < B_{\rm ext} < B_{S_{\pm},T_{\pm}} & \Rightarrow & \left\{ \begin{array}{rcc} W_{S_-,T_-} & \Rightarrow & P < 0 \\ W_{S_-,T_+} , W_{T_x,T_+} , W_{T_-,T_x} & \Rightarrow & P > 0 \end{array} \right. \label{B_rgms_c} \\
B_{S_{\pm},T_{\pm}} < B_{\rm ext} & \Rightarrow & \begin{array}{ccc} W_{S_{\pm},T_+} , W_{T_x,T_+} , W_{T_-,T_x} & \Rightarrow & P > 0 \end{array} \label{B_rgms_d}
\end{eqnarray}
\end{subequations}
where the sign of the nuclear spin polarization in each case is described in~\eref{P_sign}. Only emission spin-flip rates are shown because at low temperatures they dominate over the absorption rates, hence, they will determine the sign of nuclear spin polarization. Thus, depending on the intensity of the external magnetic field there will be different spin-flip processes dominating.

\subsection{Dependence on the hyperfine interaction intensity}

\subsubsection{Small magnetic fields} \label{rslts_smlB}

\begin{figure}
  \centering 
  \subfloat[Without feedback.]{\label{P}\includegraphics[width=0.33\textwidth]{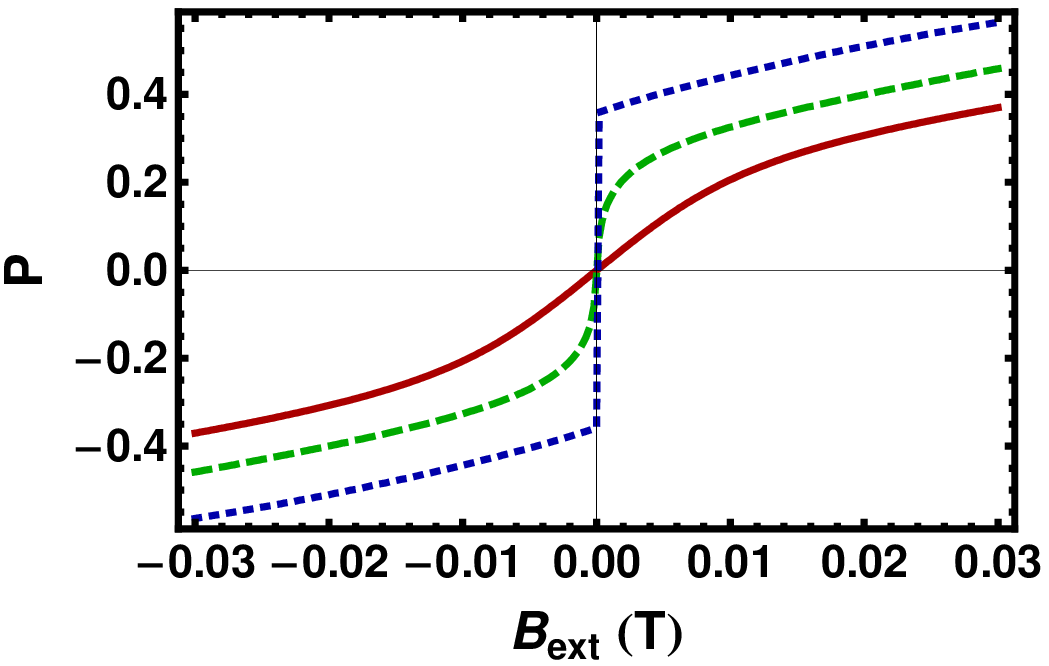}}
  \subfloat[With feedback.]{\label{Pf}\includegraphics[width=0.33\textwidth]{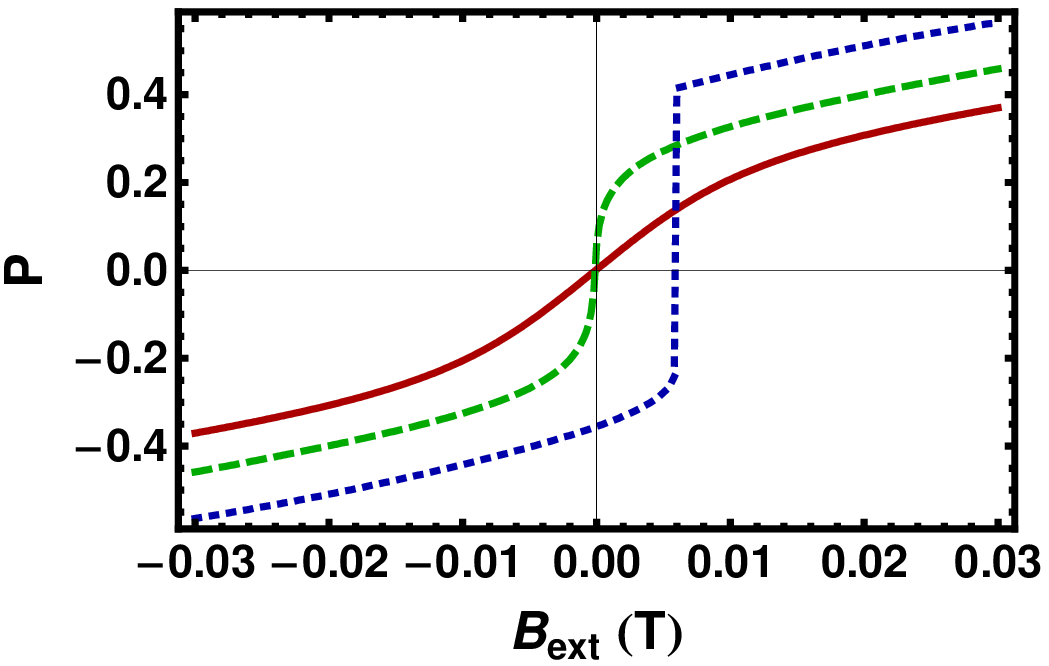}}
  \subfloat[Sweeping forwards and backwards.]{\label{Ph}\includegraphics[width=0.33\textwidth]{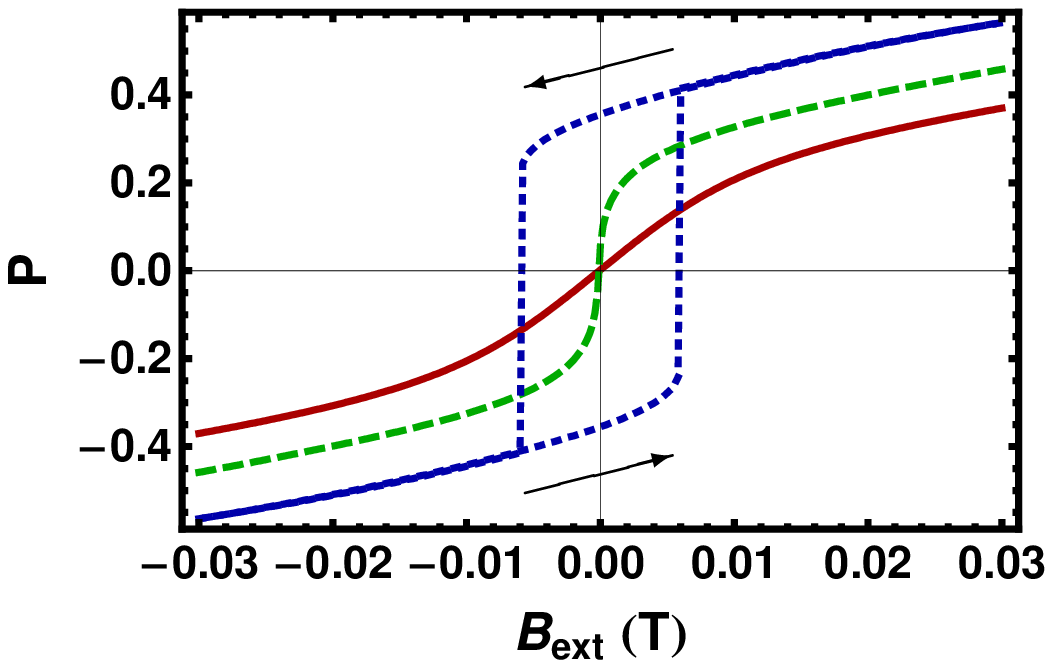}} \\
  \centering 
  \subfloat[Without feedback.]{\label{I}\includegraphics[width=0.33\textwidth]{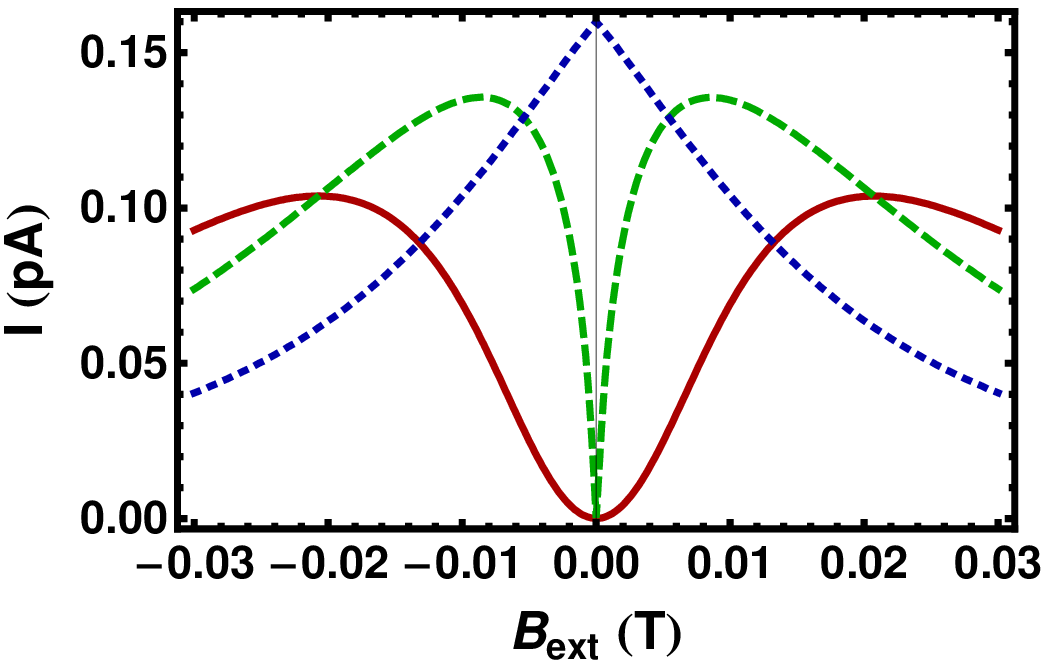}}
  \subfloat[With feedback.]{\label{If}\includegraphics[width=0.33\textwidth]{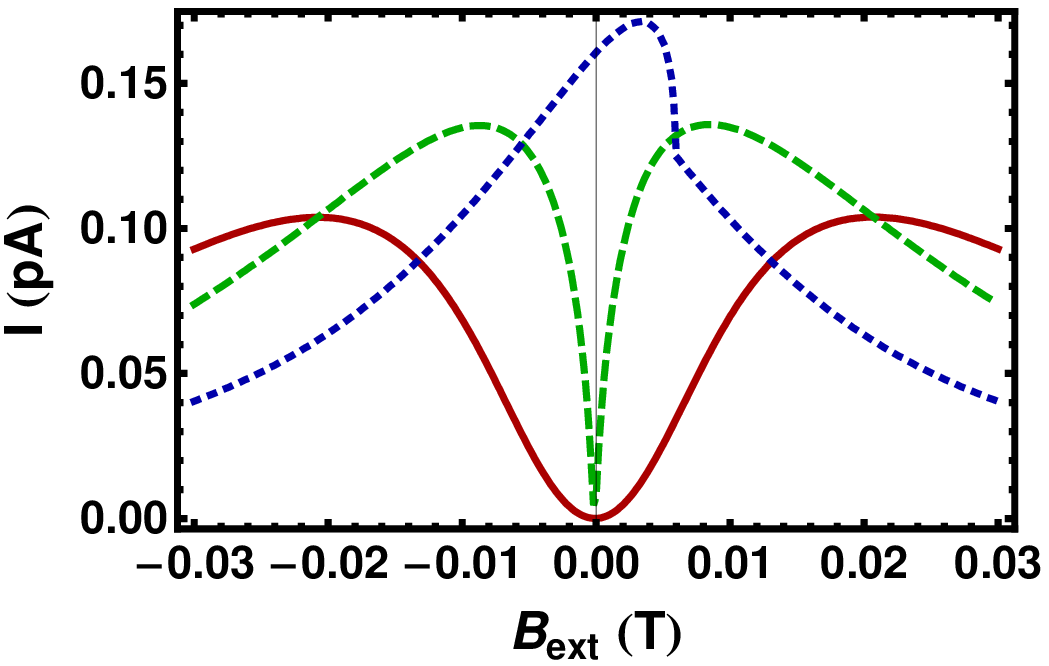}}
  \subfloat[Sweeping forwards and backwards.]{\label{Ih}\includegraphics[width=0.33\textwidth]{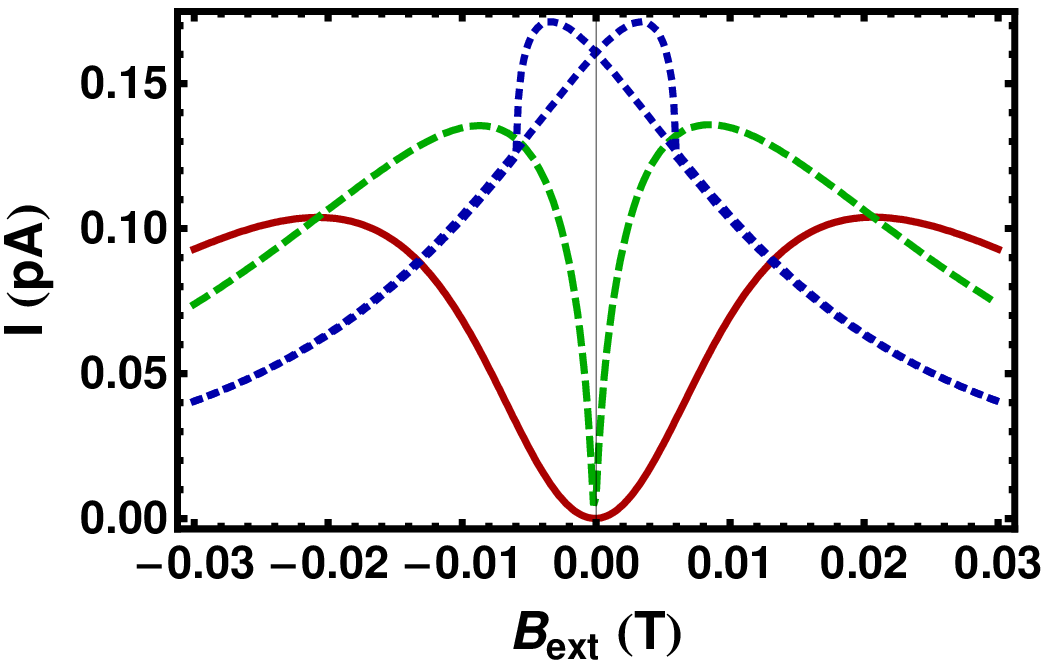}}
  \caption{a), b), c) Induced nuclear spin polarization versus external magnetic field. d), e), f) Leakage current versus external magnetic field. Parameters: $T = 120 \,$mK, $t_{LR} = 50 \, \mu$eV, $\Gamma_L = \Gamma_R = k_BT/3$, $A_R = 0.8A_L$, $\gamma = 5 \, \mu$eV, $N = 5 \times 10^4$. Initial conditions: $\rho_{1_R}(0) = 1$ and $P(0) = 0$. $A_L = 70 \, \mu$eV (solid, red online), $A_L = 80 \, \mu$eV (dashed, green online) and $A_L = 90 \, \mu$eV (dotted, blue online). Hysteresis is only observed for the largest value of the HF coupling ($A_L = 90 \, \mu$eV). The leakage current shows a dip (peak) for the smallest (largest) value of the HF coupling.}
  \label{PI}
\end{figure}

\begin{figure}
  \centering 
  \subfloat[$A_L = 70 \, \mu$eV]{\label{Eh_70}\includegraphics[width=0.35\textwidth]{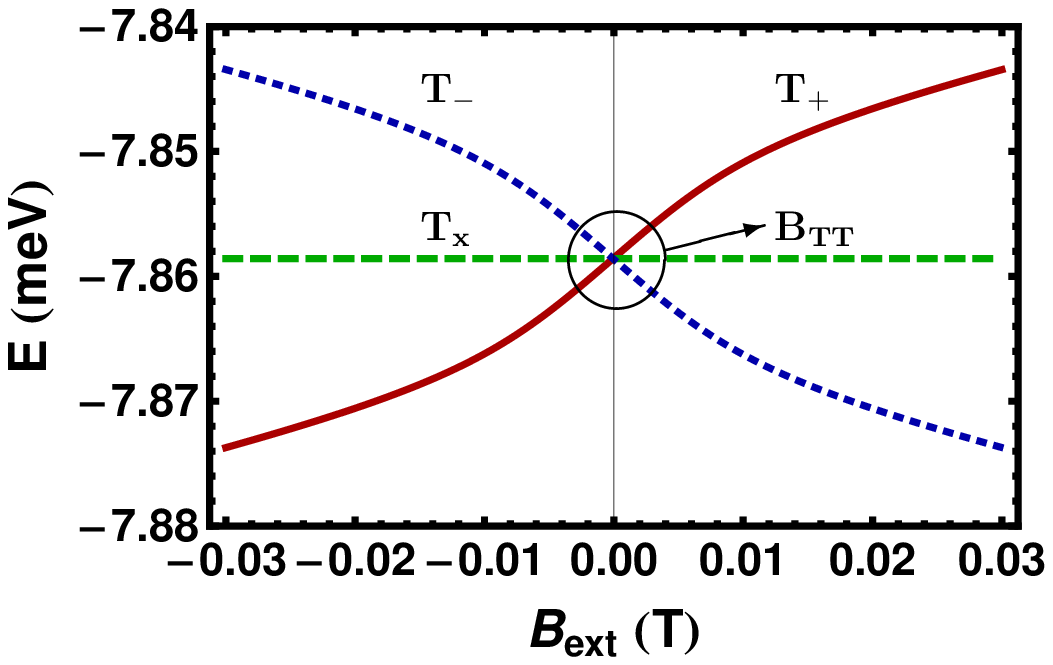}}
  \subfloat[$A_L = 90 \, \mu$eV]{\label{Eh_90}\includegraphics[width=0.35\textwidth]{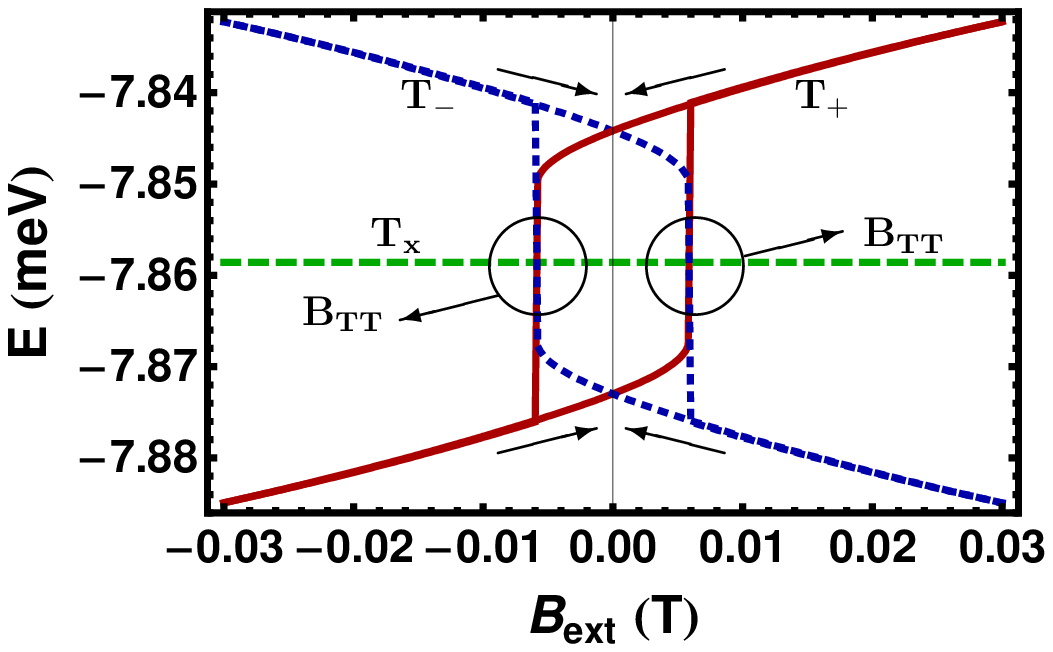}}
  \caption{Energy levels versus external magnetic field taking the spin electron-nuclei feedback into account and sweeping forwards and backwards $B_{\rm ext}$, for $A_L = 70 \, \mu$eV (a) and $A_L = 90 \, \mu$eV (b). $T_+$ (solid, red online), $T_-$ (dotted, blue online), $T_x$ (dashed, green online). Same parameters and initial conditions than in \fref{PI}. The magnetic field range considered just includes the $T_{\pm}T_x$ crossing. Singlet states are far away in energy. As in \fref{PI}, hysteresis is only observed for the largest value of the HF coupling ($A_L = 90 \, \mu$eV).}
  \label{E}
\end{figure}

In this section we consider small external magnetic field intensities. We will focus on the $B_{S_{\pm},T_{\mp}} < B_{\rm ext} < B_{S_{\pm},T_{\pm}}$ range of magnetic fields. The main spin-flip transition rates corresponding to this range of magnetic fields are shown in~\eref{B_rgms_b} and~\eref{B_rgms_c}, and the energy levels scheme is plotted in \fref{E_rgm_b} and~\fref{E_rgm_c} (see also~\fref{E}). In this case, the transitions around the $T_{\pm}T_x$ crossing give the main contribution to the current, although other possible opposite spin level transitions participate (see~\eref{B_rgms_b} and~\eref{B_rgms_c}). The $S_{\pm}T_{\mp}$ and $S_{\pm}T_{\pm}$ crossings occur for larger magnetic field intensities and will be studied in~\sref{rslts_lrgB}. \Fref{PI} shows the induced nuclear spin polarization and the leakage current through the DQD versus the external magnetic field for different values of the HF coupling ($A_L$), in this range of the external field.

The induced nuclear spin polarization versus the external field is shown in \fref{P} (without feedback), \fref{Pf} (with feedback) and \fref{Ph} (sweeping forwards and backwards with feedback). \Fref{E} shows the energy levels. When $B_{\rm ext} < B_{TT}$ ($B_{\rm ext} > B_{TT}$) the net induced nuclear spin polarization is negative (positive): Emission processes are stronger than absorption processes, hence, they mainly determine the sign of the nuclear spin polarization. In~\eref{B_rgms_b} it is shown that when $B_{\rm ext} < B_{TT}$ ($B_{\rm ext} > B_{TT}$) the dominating spin-flip rates are $W_{S_-,T_-}$, $W_{T_x,T_-}$ and $W_{T_+,T_x}$ ($W_{S_{\pm},T_+}$, $W_{T_x,T_+}$ and $W_{T_-,T_x}$), which polarize the nuclei negatively (positively). Nevertheless, in this regime absorption processes have the effect of partially compensating emission processes resulting in a finite but not complete spin polarization of the nuclei. In \fref{P}, \fref{Pf} and~\fref{Ph} it is shown that increasing the HF coupling increases the nuclei spin polarization. In this region, since $|E_{S_{\pm}} - E_{T_x}| \gg |E_{T_{\pm}} - E_{T_x}|$, the $T_{\pm} \leftrightarrow T_x$ transitions are the most important (see~\fref{E_rgm_b}, \fref{E_rgm_c} and~\fref{E}). The absolute energy difference between these levels is given by the total effective Zeeman splitting (see~\eref{egn_enrgs}):
\begin{eqnarray}
\Delta_{{\rm tot}} = \left( g\mu_B B_{{\rm ext}} + \frac{A_+}{2} \, P \right).
\end{eqnarray}
As the HF coupling intensity increases, so does the energy difference between the initial and the final states. Therefore, absorption processes become weaker with respect to the emission processes as $A_L$ increases (see~\eref{C}), allowing the nuclei spins to become more polarized (\fref{PI}).

\Fref{E} shows the energy levels of the DQD versus the external magnetic field. Electron-nuclei spin feedback is taken into account, and the external field is swept forwards and backwards. \Fref{Eh_70} shows the energy levels for the smallest HF coupling considered ($A_L = 70 \, \mu$eV). In this case: i) the induced Overhauser field is always parallel to the external field (see~\eref{Bnuc}); and ii) the $T_{\pm}T_x$ crossing occurs at $B_{\rm ext} = B_{TT} = 0 \, $T. \Fref{Eh_90} shows that for the largest value of the HF coupling ($A_L = 90 \, \mu$eV) considered here, the effect of including the feedback is to renormalize the electronic energy levels in such a way that the $T_{\pm}T_x$ crossing occurs at larger absolute magnetic field than in the previous case. When sweeping forwards (backwards) from $B_{\rm ext} < 0$ ($B_{\rm ext} > 0$) through $B_{\rm ext} = 0$, there is a negative (positive) nuclei spin polarization built up (\fref{Ph}), hence, it is still necessary to increase (decrease) the external field in order to compensate the accumulated Overhauser field and reach the $T_{\pm}T_x$ crossing. This crossing occurs now for $B_{TT} > 0$ ($B_{TT} < 0$) when sweeping forwards (backwards). Thus,~\fref{Pf} shows a small region for positive values of the external field where the Overhauser field and the external field are antiparallel. Finally,~\fref{Ph} and~\fref{Eh_90} show that, precisely in this region where the external and the induced fields are antiparallel, hysteresis is observed as the external field is swept backwards. Notice that the hysteresis is observed only for the largest value of the HF coupling intensity considered. Moreover, the size of the hysteresis loop increases with the HF coupling intensity (not shown in the figures). Summarizing: i) The amount of polarization induced in the nuclei spins depends on the competition between absorption and emission spin-flip processes; ii) the nuclear spin polarization increases with the HF coupling; and iii) for large values of the HF coupling, the nuclear spin polarization versus the external magnetic field presents a bistable region where hysteresis is observed. In this region the external field and the Overhauser field are antiparallel.

\begin{figure}
  \centering 
	\subfloat[$A_L = 70 \, \mu$eV.]{\label{rts_70}\includegraphics[width=0.3\textwidth]{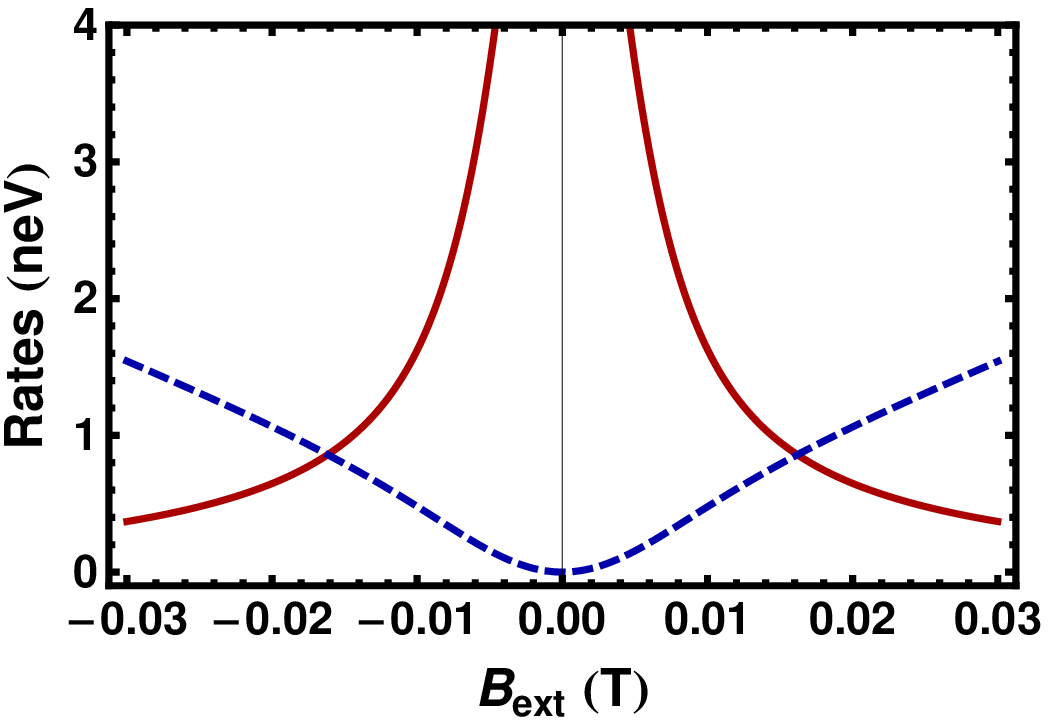}}
	\subfloat[$A_L = 80 \, \mu$eV.]{\label{rts_80}\includegraphics[width=0.3\textwidth]{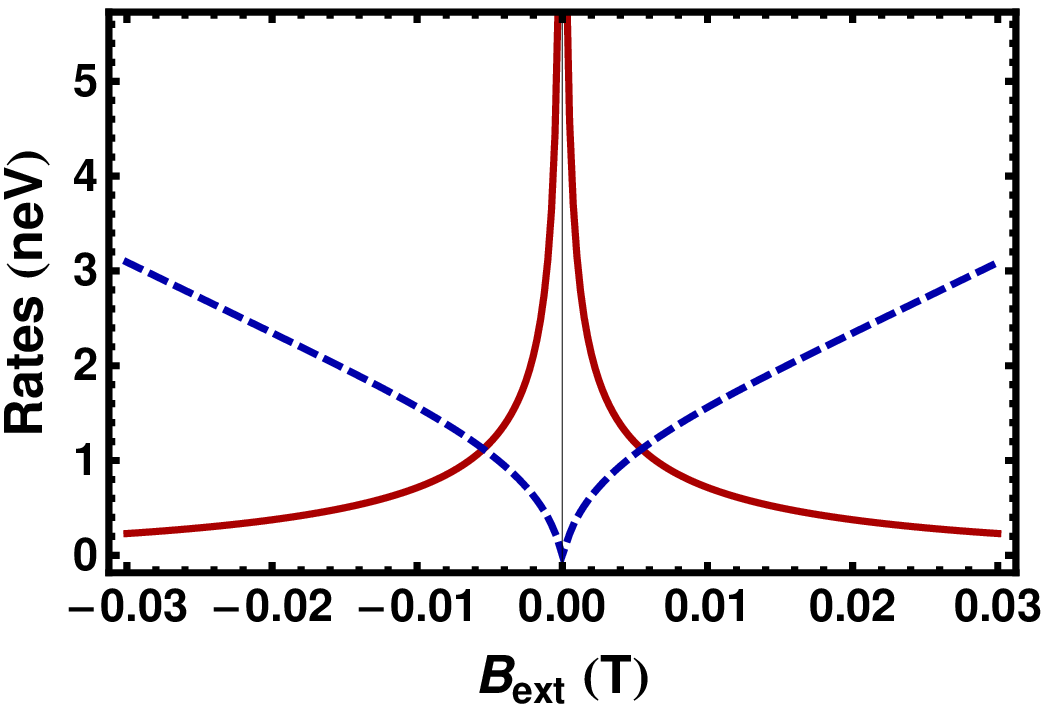}}		
	\subfloat[$A_L = 90 \, \mu$eV.]{\label{rts_90}\includegraphics[width=0.3\textwidth]{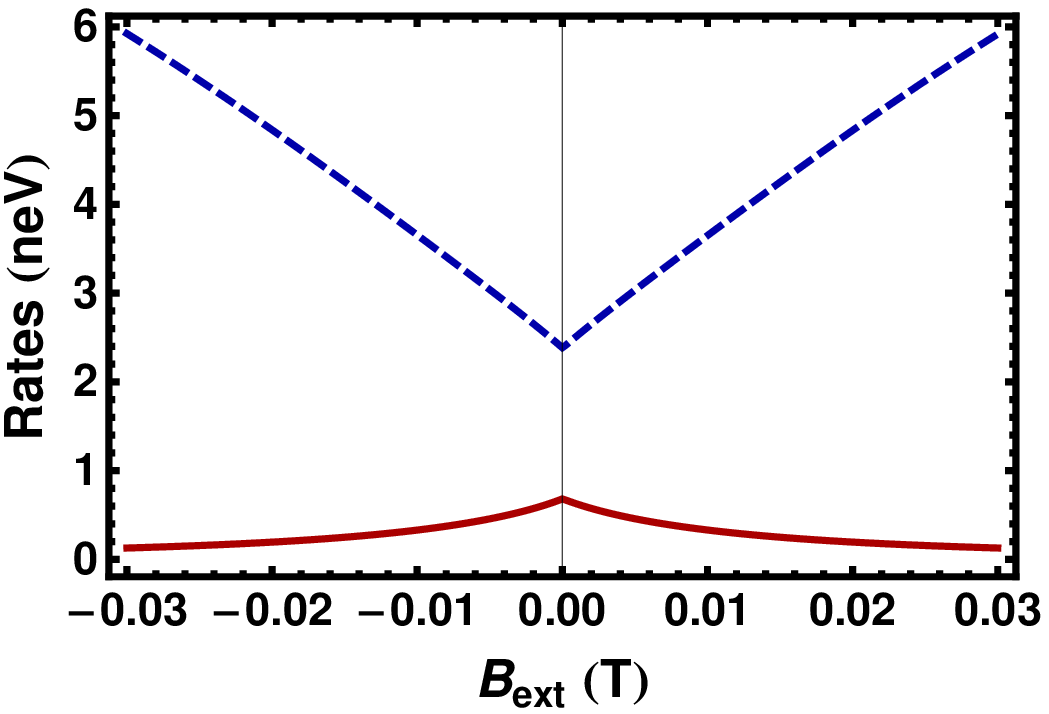}}	
  \caption{Spin-flip rates ($\omega_+^{tt}$, see~\eref{abW}; solid, red online) and tunnelling rates through contact barriers ($\beta$, see~\eref{abW}; dashed, blue online) versus external magnetic field. Same parameters and initial conditions than in \fref{PI}. In \fref{rts_70} and \fref{rts_80} there are two regions: i) $\beta > \omega_+^{tt}$, namely, electron tunnelling from $T_x$ to the right lead is {\it more} effective than electron spin-flip from $T_x$ to $T_{\pm}$ triplets (high current regime); and ii) $\beta < \omega_+^{tt}$, namely, electron tunnelling from $T_x$ to the right lead is {\it less} effective than electron spin-flip from $T_x$ to $T_{\pm}$ triplets (low current regime). In this case, the current shows a dip around zero external magnetic field (\fref{PI}). In \fref{rts_90}, $\beta > \omega_+^{tt}$ always. In this case, the current shows a peak (\fref{PI}).}
  \label{rts}
\end{figure}

We will now analyze the leakage current behaviour versus the external magnetic field. \Fref{I} (without feedback), \fref{If} (with feedback) and \fref{Ih} (sweeping forwards and backwards with feedback) shows different behaviours for the current depending on the intensity of the HF coupling. The smallest HF coupling intensity considered ($A_L = 70 \, \mu$eV) presents a current dip around the $T_{\pm}T_x$ crossing (\fref{E}), whereas the largest one ($A_L = 90 \, \mu$eV) presents a current peak. This behaviour can be understood recalling the current behaviour studied in the scheme given in~\eref{alf_bt}. Briefly, this scheme defined a low and a high current regime, comparing the $T_x \to T_{\pm}$ spin-flip rates ($W_{T_{\pm},Tx}$) with the $T_x \to |\pm\rangle$ tunnelling rates ($\Gamma_{\pm,T_x}$). These rates are shown in \fref{rts}. \Fref{rts_70} ($A_L = 70 \, \mu$eV) shows that around the $T_{\pm}T_x$ crossing $W_{T_{\pm},T_x} \gg \Gamma_{\pm,T_x}$ ($\omega_+^{tt} \gg \beta$ in diagram~\eref{alf_bt}). Therefore, it is more probable for electrons to spin-flip from $T_x \to T_{\pm}$ than to tunnel from $T_x \to |\pm \rangle$ through the contact barrier, so a current dip is observed (low current regime). By contrast, \fref{rts_90} ($A_L = 90 \, \mu$eV) shows that around the $T_{\pm}T_x$ crossing, $W_{T_{\pm},T_x} < \Gamma_{\pm,T_x}$ ($\omega_+^{tt} < \beta$ in diagram~\eref{alf_bt}). Therefore, it is more probable for electrons to tunnel from $T_x \to |\pm\rangle$ than to spin-flip from $T_x \to T_{\pm}$, so a current is enhanced and a peak is observed (high current regime). Furthermore, \Fref{P} shows that for the smallest intensity considered for the HF coupling ($A_L = 70 \, \mu$eV), the nuclear spin polarization goes to zero around the $T_{\pm}T_x$ crossing, and so does $\Gamma_{\pm,T_x}$ (see~\eref{rrts} and \fref{rts_70}), thus, a current dip is observed (\fref{I}). However, for the largest intensity considered ($A_L = 90 \, \mu$eV), the nuclear spin polarization is finite around the $T_{\pm}T_x$ crossing and $\Gamma_{\pm,T_x} > W_{T_{\pm},Tx}$ (\fref{rts_90}), thus, a current peak is observed (\fref{I}).

\Fref{I}, \fref{If} and \fref{Ih} show how the leakage current increases as the HF coupling intensity increases when the external field is close to zero. When increasing the HF coupling, the tunnelling rate $\Gamma_{\pm,T_x}$ (see~\eref{rrts}) also increases (\fref{rts}) and, therefore, so does the current. For $A_L = 70 \, \mu$eV and $A_L = 80 \, \mu$eV, a current drops to zero at $B_{\rm ext} = B_{TT} = 0$, whereas for $A_L = 90 \, \mu$eV a current peak (\fref{I}) is observed. This peak occurs at $B_{\rm ext} = B_{TT} \ne 0$ when feedback is considered (\fref{If}). Moreover, \fref{If} shows that the effect of the electron nuclei spin feedback is only appreciable for the largest value of the HF coupling ($A_L = 90 \, \mu$eV). Therefore, only in this case current hysteresis is observed when sweeping backwards the external field (\fref{Ih}).

% Rest of parametros: $g = 2$, $U = 8 \,$meV, $V_{SD} = 4 \,$meV, $\tau_{rel} = 9 \,$min

\subsubsection{Simplified model around the triplet-triplet crossing}

\begin{figure}
  \centering 
	\includegraphics[width=0.4\textwidth]{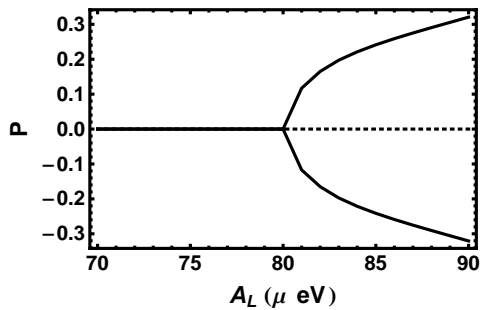}
  \caption{Solutions for~\eref{P_eff} for $B_{\rm ext} = 0$. The induced nuclei spin polarization presents a bifurcation. When $A_L \le 80.43 \, \mu$eV, there is only one stable solution, which corresponds to having the nuclei spins fully depolarized (solid line). However, when $A_L > 80.43 \, \mu$eV, $P = 0$ becomes an unstable solution (dashed line) and two stable solutions show up (solid lines). Same parameters as in \fref{PI}.}
  \label{bif}
\end{figure}

In the regime described in~\sref{rslts_smlB}, the relations $|E_{T_{\pm}} - E_{S_+}| \gg \gamma$ and $|E_{T_{\pm}} - E_{S_-}| \gg \gamma$ are well satisfied, so the rates $\omega_{\pm}^{st}$ that appear in the reduced rate equations~\eref{rdcd_eqns} can be safely neglected (see~\eref{abW} and~\eref{Wsf}) and~\eref{rdcd_eqns} becomes:
\begin{eqnarray}
\dot{\rho}_{T_x} = \omega_+^{tt} - (3\omega_+^{tt} + \beta) \rho_{T_x} - \omega_-^{tt} P \nonumber \\ \dot{P} = \omega_-^{tt} + \omega_-^{tt} \rho_{T_x} - \omega_+^{tt} P
\end{eqnarray}
Considering the stationary limit, we obtain the following equations for the stationary solutions:
\begin{subequations}
\begin{eqnarray}
\rho_{T_x} = \frac{1 - p_t^2}{3 + \frac{\beta}{\omega_+^{tt}} + p_t^2} \label{rTx_eff} \\ P = \frac{4 + \frac{\beta}{\omega_+^{tt}}}{3 + \frac{\beta}{\omega_+^{tt}} + p_t^2} \, p_t \label{P_eff}
\end{eqnarray}
\end{subequations}
where $p_t = \omega_-^{tt}/\omega_+^{tt}$. The rates $\omega_{\pm}^{tt}$ and $\beta$ are complicated functions of $P$ (see~\eref{abW}, \eref{Wsf} and~\eref{TT_rel}), thus, \eref{P_eff} cannot be solved analytically. We have obtained numerical solutions for $B_{\rm ext} = 0$ (\fref{bif}) case, although~\eref{rTx_eff} and~\eref{P_eff} hold also for finite external fields. \Fref{bif} shows that increasing the HF coupling intensity produces a {\it bifurcation} on the induced nuclei spin polarization. For $A_L \le 80.43 \, \mu$eV, the nuclei spins have {\it one} stable solution at $P = 0$, namely, the nuclei spins are fully depolarized (solid line). However, for $A_L > 80.43 \, \mu$eV, the $P = 0$ solution becomes unstable (dashed line), and {\it two} stable solutions (solid lines), with the same absolute value but with opposite signs appear. This behaviour is the same found in the full numerical solution (\fref{P}). We have found that the system undergoes the bifurcation as the slope of the linear term of the expansion around $P = 0$ of the right hand side of \eref{P_eff} is varied~\cite{Strogatz}. The slope is given by:
\begin{eqnarray} \label{slp}
s = \frac{1}{3} \left( \frac{A_+}{k_BT} - 4 \right).
\end{eqnarray}
When $s < 1$, $P = 0$ is the only stable solution, while when $s > 1$ the system presents the two stable solutions mentioned above. The bifurcation occurs at $s = 1$. Putting into~\eref{slp} the parameters for which~\fref{bif} is obtained we find that the bifurcation takes place when $A_L = 80.43 \, \mu$eV. Therefore, \eref{slp} provides an expression that relates the hyperfine coupling intensity with the temperature and the bifurcation which, in addition, can be observed through the hysteresis plots measured for the current through the DQD versus the external field~\cite{Koppens_Science_2005,Pfund_PRL_2007,Churchill_Nature_2009,Ono_PRL_2004}. Finally, we find that the agreement between the induced nuclei spin polarization obtained with the full calculation for $B_{\rm ext} = 0$, and the one obtained with this simplified model is very good. 

From the current through the right contact barrier $I_R = (\Gamma_{+,T_x} + \Gamma_{-,T_x}) \rho_{T_x}$ (see~\eref{crnt}), and from~\eref{rTx_eff} we obtain the following expression for the current through the DQD:
\begin{eqnarray} \label{crnt_eff}
I = \left( 1 + \frac{1}{2 \mathcal{N}^2} \right) \frac{1 - p_t^2}{3 + \frac{\beta}{\omega_+^{tt}} + p_t^2} \, \beta
\end{eqnarray}
Notice that $\beta = 0$ when $P = 0$ (see~\eref{abW}), thus, the current will be zero when the nuclei spins are fully depolarized. In the previous discussion, we have shown that the induced nuclei spin polarization shows a bifurcation when increasing the HF coupling intensity. Therefore, in the range of $A_L$ where the nuclei spins are fully depolarized (\fref{bif}) no current flows through the DQD at $B_{\rm ext} = 0$. However, in the range of $A_L$ where the nuclei have a non zero spin polarization (\fref{bif}), a finite current flows through the DQD at $B_{\rm ext} = 0$, since $I (B_{\rm ext} = 0) \propto P^2$ to lowest order in $P$. Furthermore, \eref{crnt_eff} shows that the current depends on the ratio $\beta/\omega_+^{tt}$. These rates are shown in~\fref{rts}, and motivate the physical picture we have used to understand the transition from a current dip to a current peak. Therefore, this simplified model shows that the bifurcation obtained for the induced nuclei spin polarization together with the ratio $\beta/\omega_+^{tt}$, describe the transition from a current dip to a current peak (\fref{I}).

\subsubsection{Large magnetic fields} \label{rslts_lrgB}

\begin{figure}
  \centering
  \subfloat[$A_L = 70 \, \mu$eV]{\label{E3h_70}\includegraphics[width=0.35\textwidth]{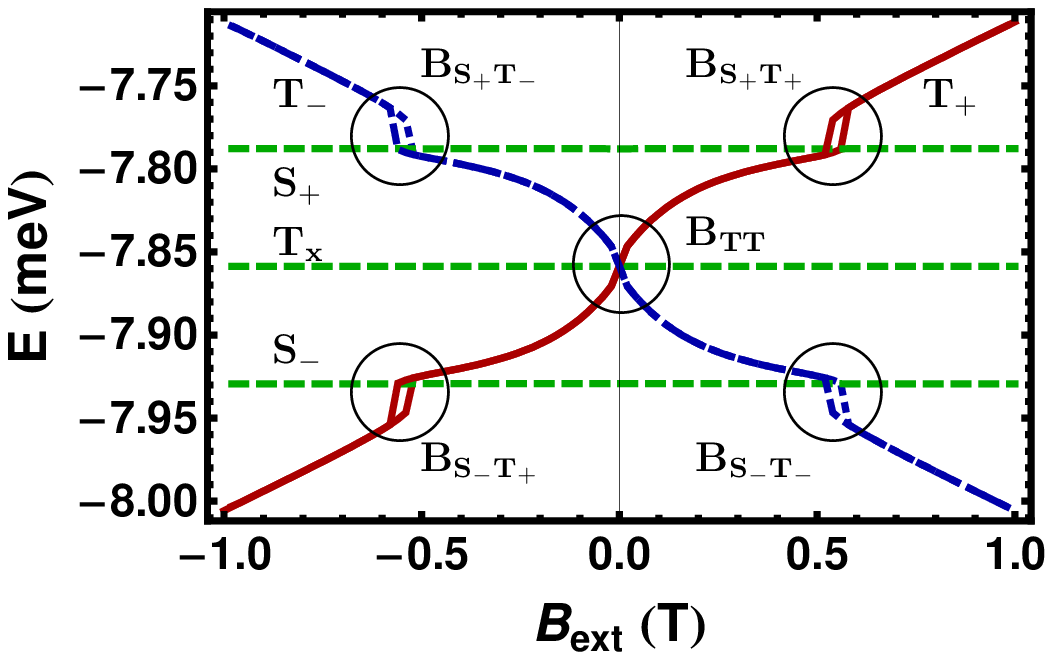}}
  \subfloat[$A_L = 90 \, \mu$eV]{\label{E3h_90}\includegraphics[width=0.35\textwidth]{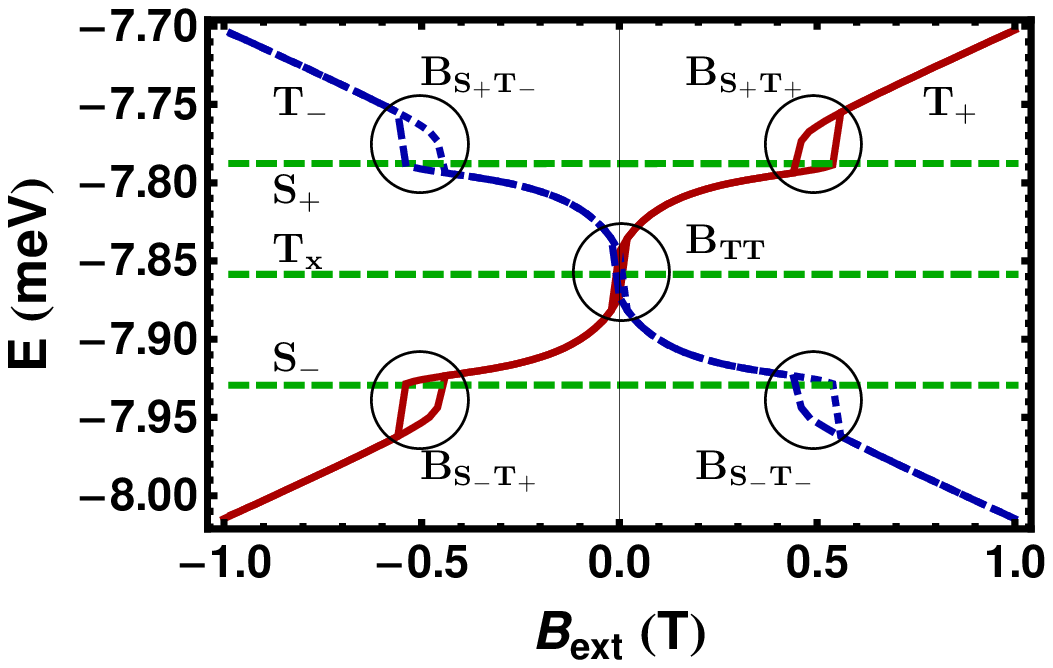}} \\
  \centering 
  \subfloat[$A_L = 70 \, \mu$eV]{\label{P3h_70}\includegraphics[width=0.35\textwidth]{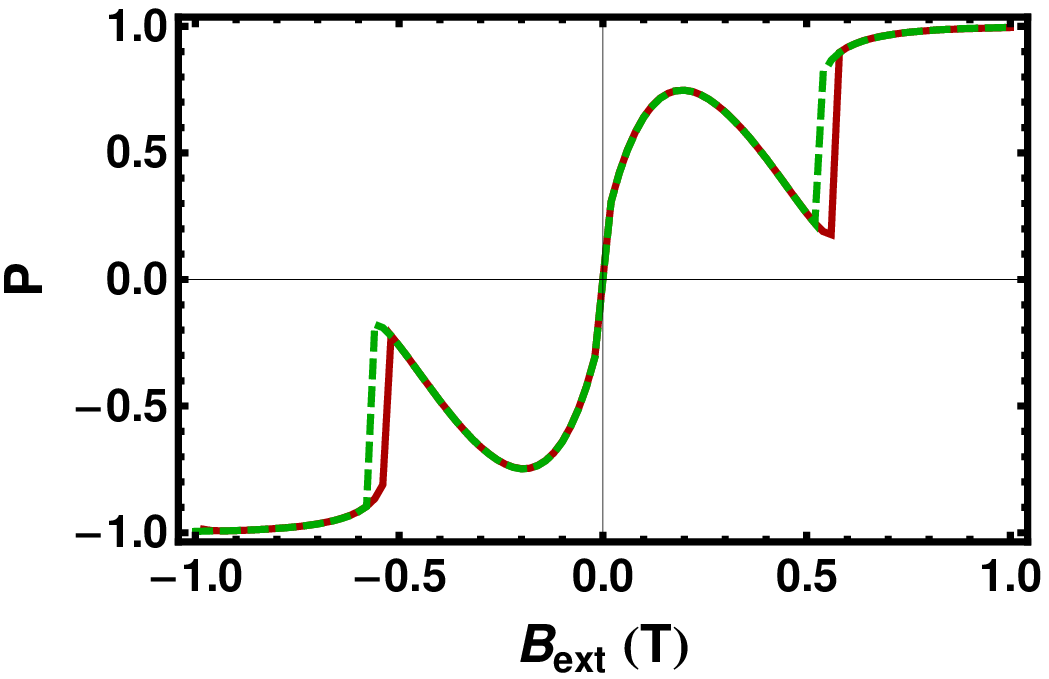}}
  \subfloat[$A_L = 90 \, \mu$eV]{\label{P3h_90}\includegraphics[width=0.35\textwidth]{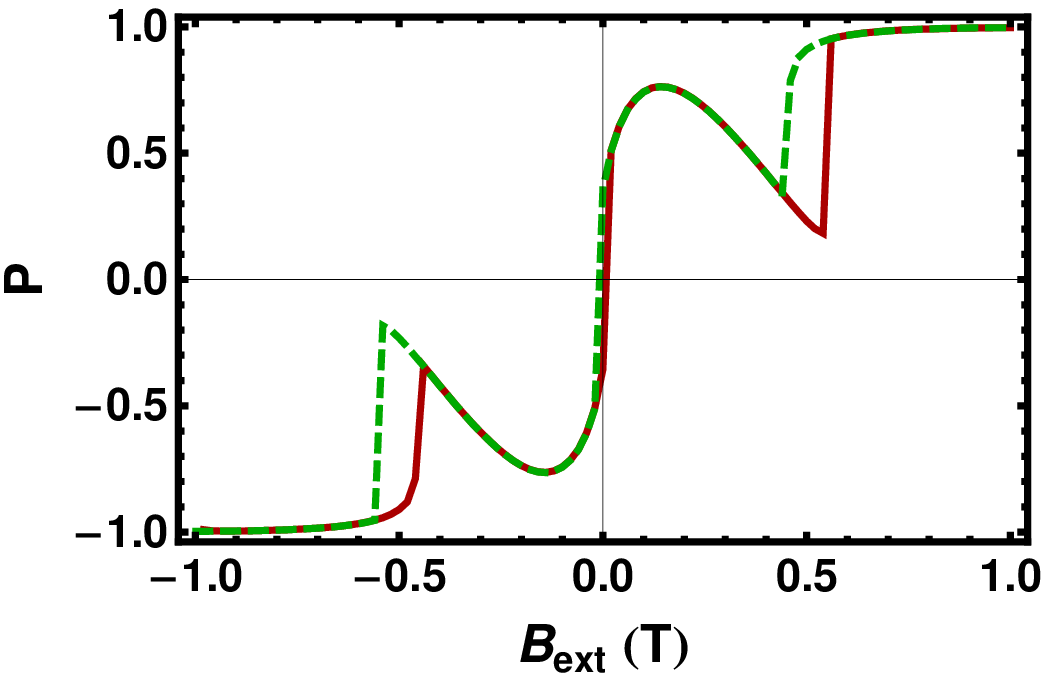}} \\
  \centering 
  \subfloat[$A_L = 70 \, \mu$eV]{\label{I3h_70}\includegraphics[width=0.35\textwidth]{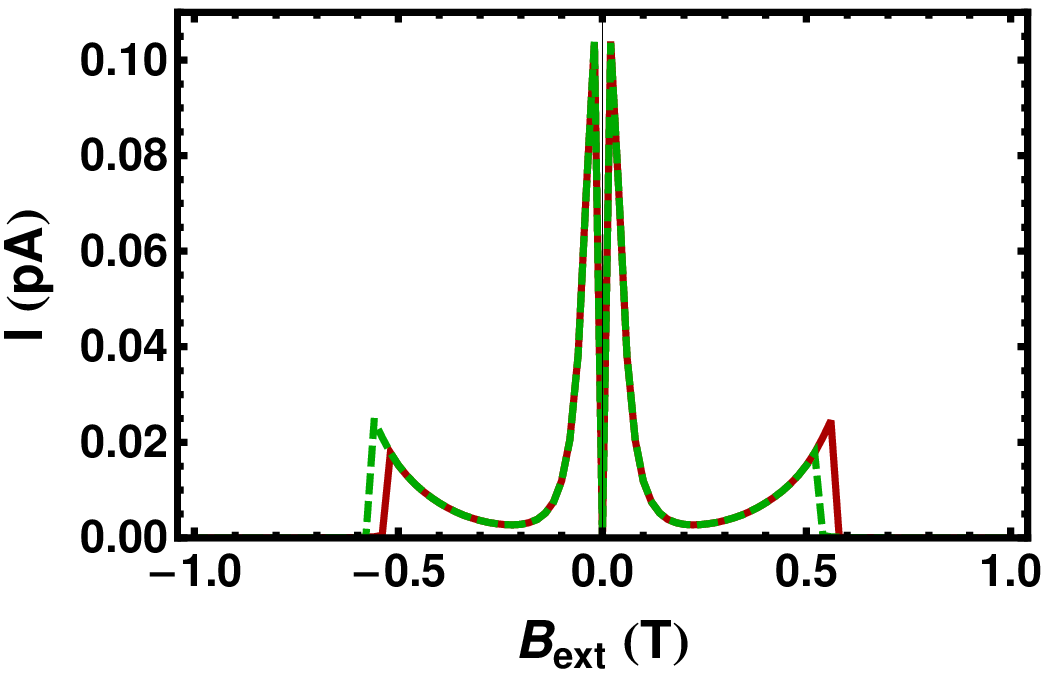}}
  \subfloat[$A_L = 90 \, \mu$eV]{\label{I3h_90}\includegraphics[width=0.35\textwidth]{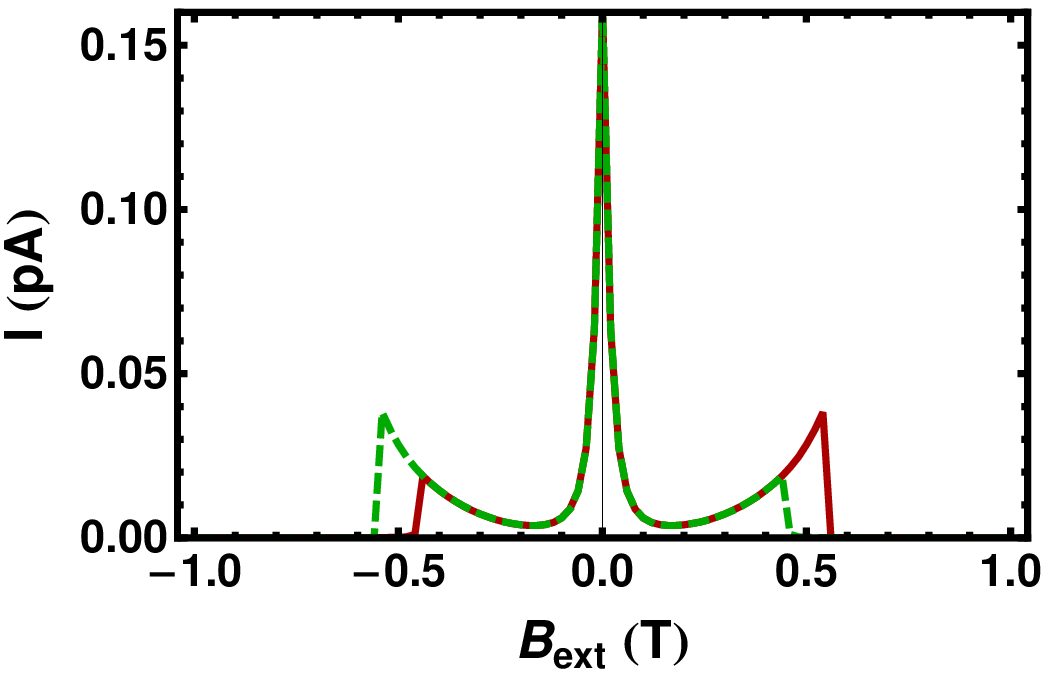}}
  \caption{a), b) Energy levels versus external magnetic field. $T_+$ (solid, red online), $T_-$ (dotted, blue online), $T_x$, $S_+$, $S_-$ (dashed, green online); c), d) induced nuclear spin polarization, and e), f) leakage current versus external magnetic field sweeping forwards (solid, red online) and backwards (dashed, green online). Same parameters and initial conditions than in \fref{PI}. The range of magnetic fields considered includes both ST and the $T_{\pm} T_x$ crossings. Hysteresis is observed for both values of the HF coupling intensity at the ST crossings. Recall that in \fref{PI}, where only the $T_{\pm}$ crossing was shown, hysteresis was only found for the largest value of the HF intensity ($A_L = 90 \, \mu$eV). The current shows now three peaks corresponding mainly to each of the levels crossings.}
  \label{PIE3}
\end{figure}

In this section, we consider larger magnetic fields than in the previous case in order to account for the ST crossings. This means sweeping the external field from $B_{\rm ext} < B_{S_{\pm},T_{\mp}}$ to $B_{\rm ext} > B_{S_{\pm},T_{\pm}}$ (see schematic~\fref{E_rgm_a} and~\fref{E_rgm_d}). Therefore all three level crossings ($S_{\pm}T_{\mp}$, $S_{\pm}T_{\pm}$ and $T_{\pm}T_x$) will be considered (see~\fref{E3h_70} and~\fref{E3h_90}). \Fref{PIE3} shows the DQD energy levels, the induced nuclear spin polarization and the leakage current through the DQD versus the external magnetic field, for $A_L = 70 \, \mu$eV and $A_L = 90 \, \mu$eV. In this case all figures include feedback and the external field is swept forwards and backwards.

Nuclear spin polarization versus the external field is shown in \fref{P3h_70} ($A_L = 70 \, \mu$eV) and \fref{P3h_90} ($A_L = 90 \, \mu$eV). The external field sweeping starts for $B_{\rm ext} < B_{S_{\pm},T_{\mp}}$. Initially, the nuclei spins are completely depolarized. The DQD energy level distribution shows that $T_+$ triplet is the ground state (see~\fref{E3h_70} and~\fref{E3h_90}). In this case, all spin-flip emission rates polarize negatively the nuclei spins (see~\eref{B_rgms_a}) and, hence, they become almost completely negatively polarized ($P \simeq -1$). Furthermore, since the probability of finding a nuclei with spin $+1/2$ is nearly zero when $P \sim -1$ (see~\eref{Wm}), all emission rates become approximately zero (see~\eref{Wsf}), and the nuclei spin polarization remains roughly constant until $B_{\rm ext}$ reaches the $S_{\pm}T_{\mp}$ crossings. Only close enough to the $S_{\pm}T_{\mp}$ level crossings, spin-flip absorption processes become significant and the nuclei spins dramatically depolarize (\fref{P3h_70} and~\fref{P3h_90}). In the $B_{S_{\pm},T_{\mp}} < B_{\rm ext} < B_{S_{\pm},T_{\pm}}$ range of magnetic fields (schematic~\fref{E_rgm_b} and~\fref{E_rgm_c}), several spin-flip rates compete in order to polarize the nuclei spins in opposite directions (see~\eref{B_rgms_b} and~\eref{B_rgms_c}). When $B_{S_{\pm},T_{\mp}} < B_{\rm ext} < B_{TT}$ (scheme in~\fref{E_rgm_d}), most of the spin-flip emission rates polarize negatively the nuclei spins (see~\eref{B_rgms_b}), and the resulting nuclear spin polarization is negative in this region. However, as the levels approach the $T_{\pm}T_x$ crossings, spin-flip absorption processes become more relevant, and the nuclei spins slowly depolarize. When $B_{TT} < B_{\rm ext} < B_{S_{\pm},T_{\pm}}$ (scheme in~\fref{E_rgm_c}), most of the spin-flip emission rates polarize positively the nuclei spins (see~\eref{B_rgms_c}), and the nuclear spin polarization is positive in this region. Finally, when $B_{\rm ext} > B_{S_{\pm},T_{\pm}}$, the $T_-$ triplet is the ground state (scheme in~\fref{E_rgm_d}), and~\eref{B_rgms_d} shows that all spin-flip emission rates polarize positively the nuclei spins, thus, they become almost completely positively polarized ($P \simeq 1$). The probability of finding a nuclei with spin $-1/2$ is nearly zero as $P \sim 1$ (see~\eref{Wm}), thus, all emission rates become approximately zero (see~\eref{Wsf}), and the nuclei spin polarization remains roughly constant.

\Fref{P3h_70} and~\fref{P3h_90} show that the feedback between electron and nuclei spins produces hysteresis in the nuclear polarization around the ST-crossings. Recall that for small magnetic fields (\sref{rslts_smlB}), where only $T_{\pm}T_x$ crossings participate, hysteresis was not observed for small HF intensities, but only the largest HF coupling intensity considered here ($A_L = 90 \, \mu$eV). For larger external magnetic fields, however, hysteresis shows up at ST crossings even for the smallest HF intensity considered ($A_L = 70 \, \mu$eV). Finally, as for small magnetic fields, hysteresis is larger as the HF coupling increases.

The leakage current through the DQD versus the external field is shown in \fref{I3h_70} ($A_L = 70 \, \mu$eV) and \fref{I3h_90} ($A_L = 90 \, \mu$eV). The current presents three peaks, each of them corresponding to one of the three possible level crossings. When $B_{\rm ext} < B_{S_{\pm}T_{\mp}}$ ($B_{\rm ext} > B_{S_{\pm}T_{\pm}}$), the current is zero because electrons are trapped in the $T_+$ ($T_-$) triplet state (\fref{E3h_70} and \fref{E3h_90}), thus, in these ranges of the external field spin-flip emission rates are nearly zero. In between these crossings (when $B_{S_{\pm},T_{\mp}} < B_{\rm ext} < B_{S_{\pm},T_{\pm}}$), current is strongly quenched but nevertheless finite. This case has been discussed in~\sref{rslts_smlB}. Finally, the feedback between the nuclei and the electron spins also produces hysteresis in the current around the ST crossings, when sweeping the external field forwards and backwards.

\subsection{Dependence on the interdot tunnelling strength} \label{rslts_tnl}

\begin{figure}
  \centering 
  \subfloat[$t_{LR} = 50 \, \mu$eV.]{\label{ETf_0_80}\includegraphics[width=0.35\textwidth]{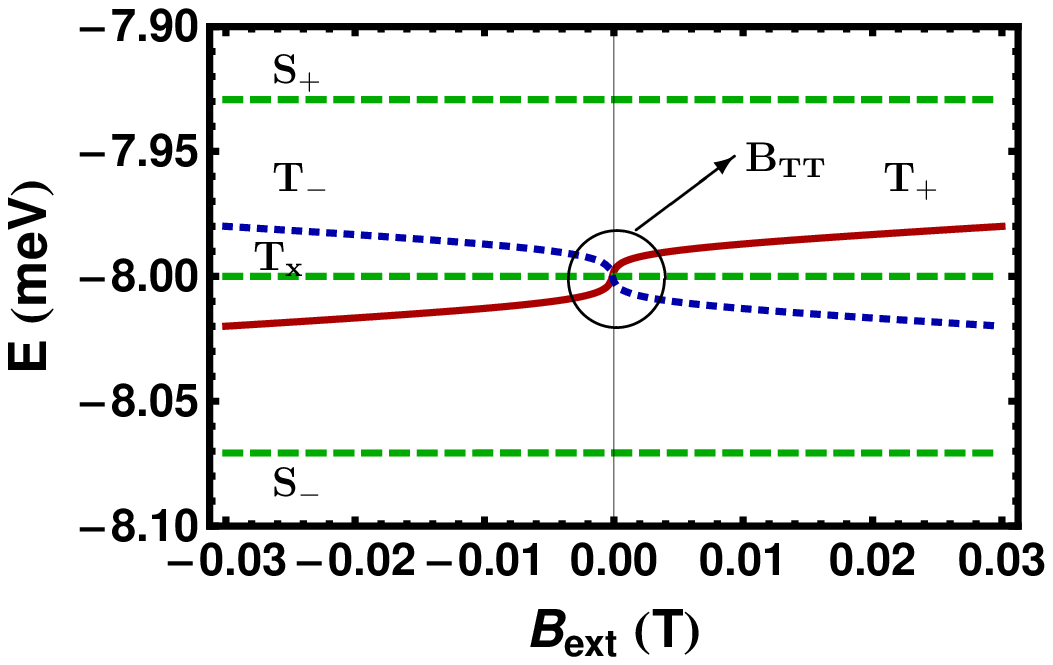}}
  \subfloat[$t_{LR} = 0.01 \, \mu$eV.]{\label{ETf_3_80}\includegraphics[width=0.35\textwidth]{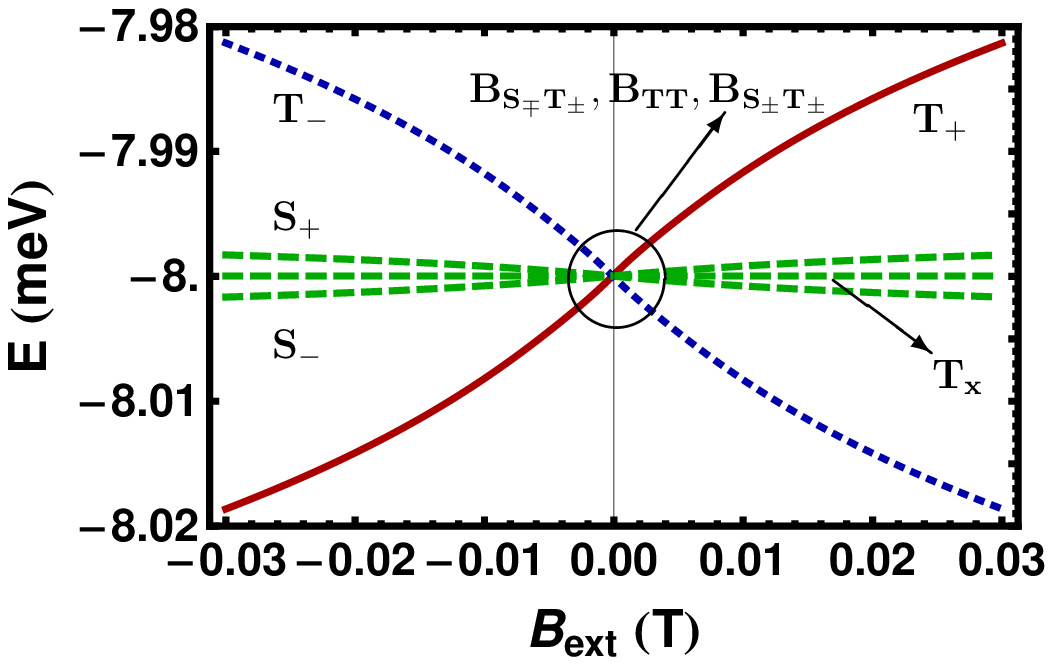}} \\
  \centering 
  \subfloat[$t_{LR} = 50 \, \mu$eV.]{\label{PTf_50}\includegraphics[width=0.35\textwidth]{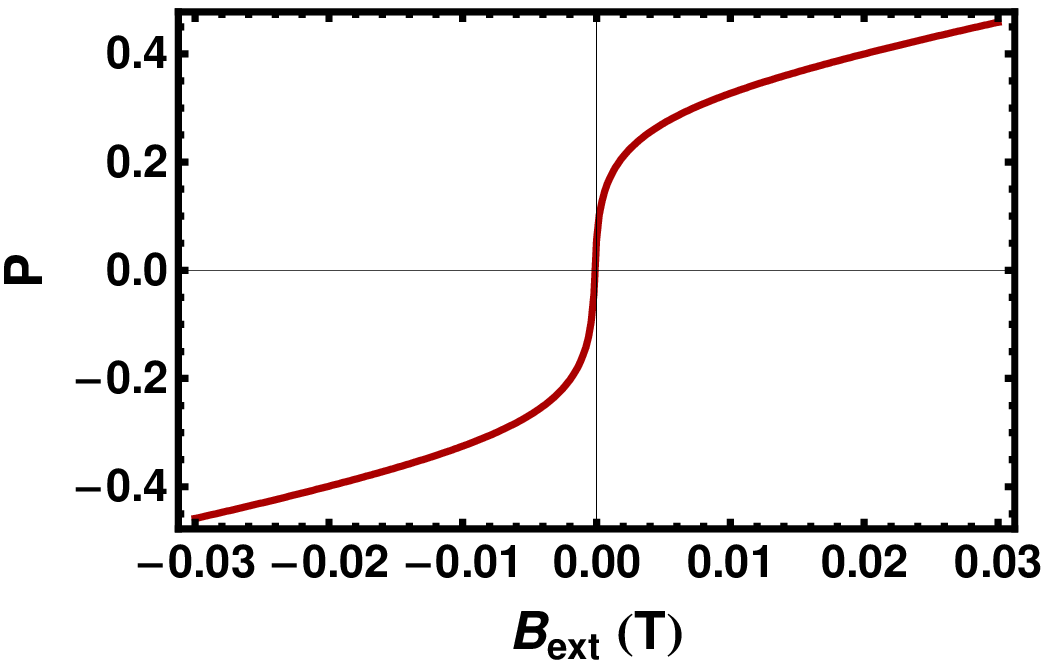}}
  \subfloat[$t_{LR} = 0.01 \, \mu$eV.]{\label{PTf_.01}\includegraphics[width=0.35\textwidth]{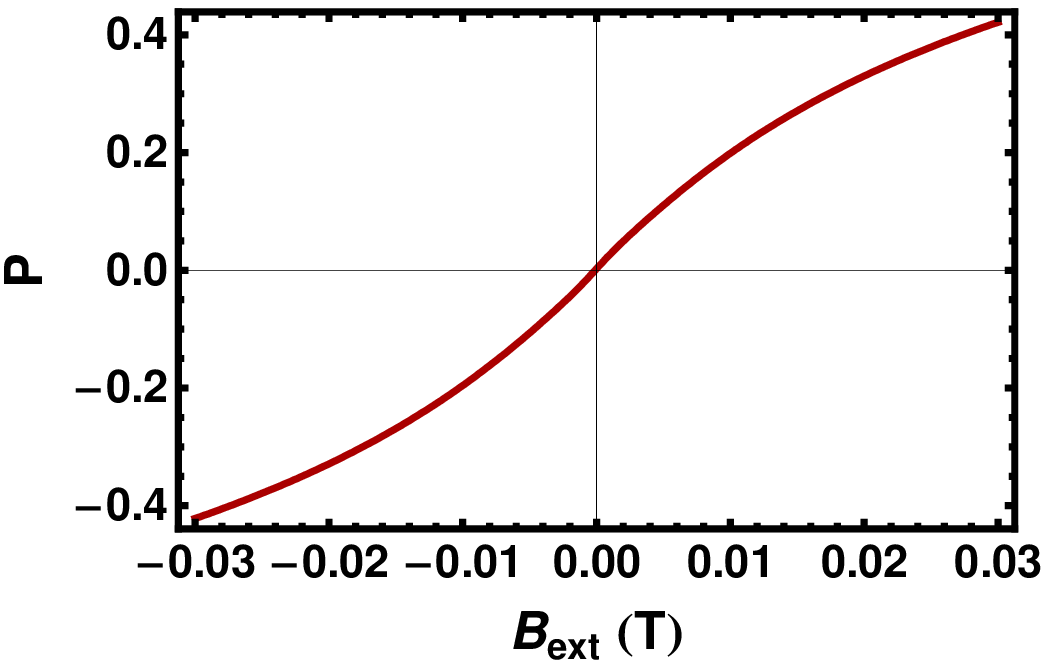}} \\
  \centering 
  \subfloat[$t_{LR} = 50 \, \mu$eV.]{\label{ITf_50}\includegraphics[width=0.35\textwidth]{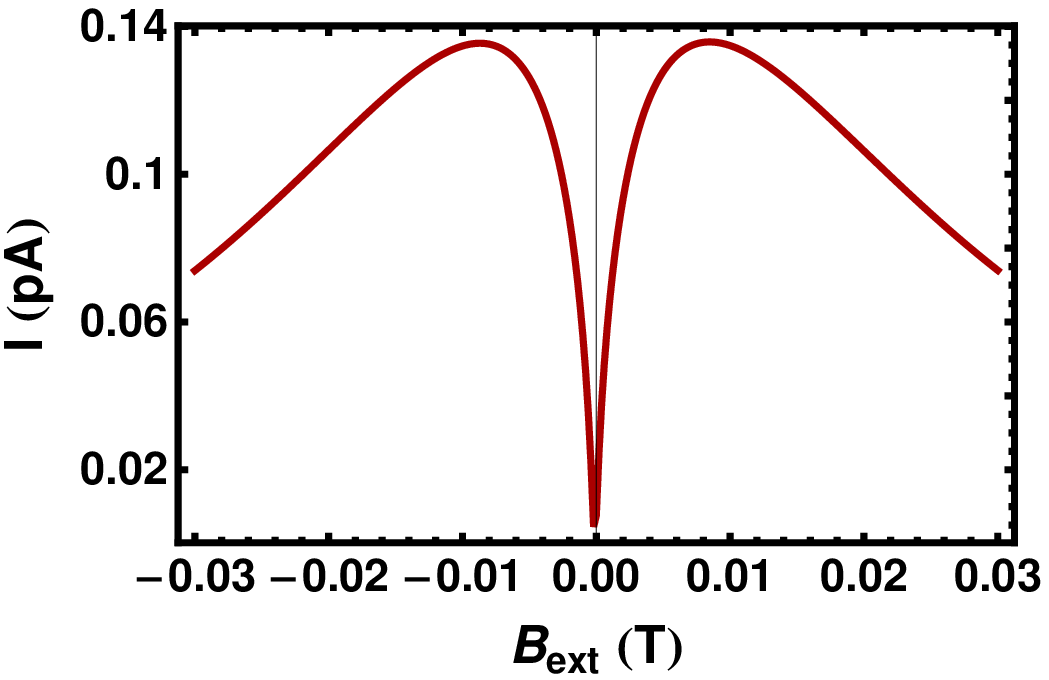}}
  \subfloat[$t_{LR} = 0.01 \, \mu$eV.]{\label{ITf_.01}\includegraphics[width=0.33\textwidth]{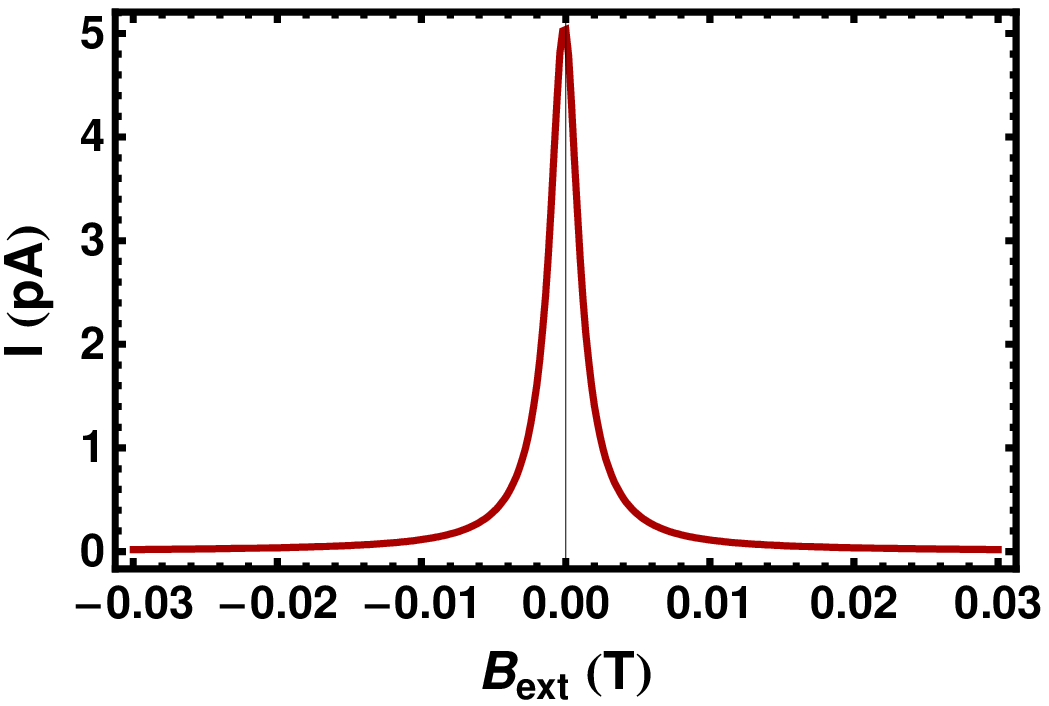}}
  \caption{a), b) Energy levels versus external magnetic field. $T_+$ (solid, red online), $T_-$ (dotted, blue online), $T_x$, $S_+$, $S_-$ (dashed, green online); c), d) induced nuclear spin polarization, and e), f) leakage current versus external magnetic field. $A_L = 80 \, \mu$eV. The remaining parameters and the initial conditions are the same as in \fref{PI}. The current shows a dip (peak) for the largest (smallest) value of the interdot tunnel intensity. Although feedback between the electron and the nuclear spin polarization is taken into account in these figures, hysteresis is not observed for this set of parameters.}
  \label{PIET}
\end{figure}

In this section, we show the leakage current and the induced nuclear spin polarization dependence on the interdot tunnelling intensity. Interdot tunnel varies a lot from one experiment to another and can be externally tuned. For instance, in~\cite{Ono_Science_2002} it is estimated to be around $30 \, \mu$eV, while in~\cite{Koppens_Science_2005} around $0.2 \, \mu$eV. This justifies the large difference between the two values that we have chosen. We consider small external magnetic fields, and two different interdot tunnel values (\fref{ETf_0_80} and \fref{ETf_3_80}). For the largest value of the interdot tunnel ($t_{LR} = 50 \, \mu$eV) only the $T_{\pm}T_x$ crossing participate in the current. However, for the smallest value of the interdot tunnel ($t_{LR} = 0.01 \, \mu$eV), all crossings: $S_{\pm},T_{\mp}$, $T_{\pm}T_x$ and $S_{\pm},T_{\pm}$ participate in the current, for the same range of the external field. \Fref{PIET} shows also the induced nuclear spin polarization (\fref{PTf_50} and \fref{PTf_.01}) and the leakage current through the DQD (\fref{ITf_50} and \fref{ITf_.01}) versus the external magnetic field for the two interdot tunnelling intensities chosen: $t_{LR} = 50 \, \mu$eV and $t_{LR} = 0.01 \, \mu$eV, and for $A_L = 80 \, \mu$eV.

\Fref{PTf_50} and \fref{PTf_.01} show that for the smallest interdot tunnelling ($t_{LR} = 0.01 \, \mu$eV), the behaviour of the polarization versus the external magnetic field is smoother than for the largest interdot tunnelling considered ($t_{LR} = 50 \, \mu$eV), due to the stronger competition between the energy absorption and emission processes in the former case. For the small interdot tunnelling, absorption is more efficient than for the large interdot tunnelling value.

\Fref{ITf_50} and \fref{ITf_.01} show that decreasing the interdot tunnelling, the current versus the magnetic field presents again a transition from a dip to a peak. As in the previous case, where we discussed the current behaviour as a function of the intensity of the HF interaction, a dip or a peak features observed in the current can be understood comparing the spin-flip rate between $T_{\pm}$ and $T_x$ and the tunnelling rate through the barrier contact between $T_x$ and $|\pm \rangle$ states (\fref{rts_80}). However, unlike in the previous case, where the HF coupling had to be increased to observe the transition, now, in order to go from a dip to a peak, the interdot tunnelling must decrease.

%% file: conclusion.tex
\section{Conclusions} \label{clns}

We have studied the leakage current through a coherently coupled DQD in SB regime. Spin relaxation due to HF interaction between the spins of the electrons in the DQD and the nuclei spins lifts SB producing leakage current. Moreover, the spin interaction between electrons tunnelling through the DQD and nuclei, induces dynamical nuclear spin polarization that is in general non-negligible. We have investigated the behaviour of both the induced nuclear spin polarization and the leakage current as a function of an external magnetic field. Our three main results are: i) the leakage current shows a dip or a peak depending on the intensities of both the HF interaction and the interdot tunnel strength. We have shown that for large (small) HF coupling (interdot tunnelling) intensities the current shows a peak. On the contrary, for small (large) HF coupling (interdot tunnelling) intensities the current shows a dip. Large (small) HF couplings (interdot tunnelling) indicate strong mixing between the singlet-triplet subspaces, namely, the effective Zeeman splitting difference between the dots is non-negligible with respect to the exchange energy. In this case the leakage current shows a peak. On the contrary, small (large) HF couplings (interdot tunnelling) indicate a weak mixing between the singlet-triplet subspaces, the Zeeman splitting difference between dots becomes negligible with respect to the exchange energy, and the system is mostly blocked in the triplet subspace. In this case the leakage current shows a dip. The crossover from a dip to a peak is, thus, obtained by increasing (decreasing) the HF interaction (interdot tunnelling strength). ii) For a wide external magnetic field sweeping range, we have shown that the leakage current shows three main peaks. Two satellite peaks corresponding to the ST crossings and a central peak corresponding to the triplets crossing. iii) We have observed hysteresis in both the leakage current and the induced nuclear spin polarization as a function of an external magnetic field. This hysteretic behaviour is a consequence of the dynamical nuclear spin polarization interacting with the electron spin that tunnel through the DQD structure. Finally, we have shown that the size of the hysteresis region strongly depends on the HF interaction intensity and the interdot tunnel strength. 

Our results are a contribution to ongoing efforts at understanding and controlling spin relaxation in qubits, which limitate their performance for quantum information and quantum computation purposes.